\documentclass[11pt]{article}

\usepackage[a4paper,margin=1in]{geometry}

\usepackage{amsmath,amssymb}
\usepackage{booktabs}
\usepackage{graphicx}
\usepackage{hyperref}
\hypersetup{hidelinks}
\usepackage{setspace}
\usepackage{float}
\usepackage{rotating}
\usepackage{subcaption}

\newcommand{\autolab}{AutoLab}

\title{AutoLab: An Internet-Accessible Experimental Platform for
Operational World Models in Wireless Networks}

\author{Jiunn-Tsair Chen, Jia-Shung Wang,\\
Chi-Yun Hsieh, and Jack Shi Jie Luo\\[0.4em]
WNC Corporation\\
\href{mailto:jt.chen@wnc.com.tw}{\texttt{jt.chen@wnc.com.tw}},
\href{mailto:JS.Wang@wnc.com.tw}{\texttt{JS.Wang@wnc.com.tw}}\\
\href{mailto:Jiyun.Hsieh@wnc.com.tw}{\texttt{Jiyun.Hsieh@wnc.com.tw}},
\href{mailto:Jack.Luo@wnc.com.tw}{\texttt{Jack.Luo@wnc.com.tw}}}

\date{}

\begin{document}
\maketitle

\begin{abstract}
Operational World Models (OWMs) require structured interaction with the
physical world: they must observe operational state, impose controlled
actions, measure consequences, preserve experience, and use that experience
to support prediction and preventive decision making.  This paper presents
\autolab{}, an Internet-accessible experimental platform that provides these
physical grounding functions for wireless-network OWMs.  A remote researcher
can inspect a live test site, reconstruct recent state history, author a
time-ordered experiment containing robot, access-point, wireless-scan, and
traffic operations, validate it locally, and submit the same explicit contract
for registration.  At the physical site, a centralized network management
system (NMS) revalidates and schedules the experiment, coordinates distributed
device agents, estimates robot location from raw camera observations, supervises
physical execution, and records measurements and diagnostic state.  The same
Python validation rules are used at the authoring and registration boundaries,
while destination-level commands isolate Internet users from low-level device
control.  An end-to-end mobility trace demonstrates how the NMS detects a large
physical movement error, issues corrective motion, and restores the intended
experiment trajectory without exposing that error to the script writer or
robot agent.  \autolab{} therefore contributes a reproducible
Internet-to-physical experimentation and feedback infrastructure, rather than
claiming a completed autonomous OWM.  Its structured spatial--temporal records
provide the experience needed for continuing work on generative scenarios,
predictive risk evaluation, and preventive wireless-network management.
\end{abstract}


\section{Introduction}
\label{sec:introduction}

Wireless-network research increasingly requires controlled interaction with
physical environments.  Simulation and offline data remain valuable, but they
cannot alone expose all effects produced by real propagation, interference,
traffic, mobility, device behavior, and implementation uncertainty.  A useful
experimental system must therefore do more than reproduce a physical setting
digitally: it must support observation, intervention, measurement, history,
and repeated interaction with the physical network.

The work reported here was not retrofitted to a world-model framework.
Its original objective has been to use (i) accumulated experience and
(ii) currently observed scenario states to predict how wireless operating
conditions may evolve, and then to interact with those predicted future
scenarios.  This predictive and preventive objective is more accurately
described by an \emph{Operational World Model} (OWM) than by the commonly
perceived replication-centered meaning of a Digital Twin.

\autolab{} is a major step in the continuing construction of this OWM.  It
turns a physical wireless site into an Internet-accessible, programmable,
observable, and repeatable experimental environment.  The complete autonomous
OWM is still under development; this paper positions and demonstrates
\autolab{} as its physical observation, intervention, execution, and feedback
infrastructure.

The contributions of this paper are:
\begin{enumerate}
  \item an OWM-oriented physical experimentation platform for wireless
        networks;
  \item an Internet-to-physical lifecycle integrating experiment creation,
        validation, registration, scheduling, execution, observation, history,
        and result retrieval;
  \item a reproducible experiment specification based on explicit initial
        conditions and time-ordered robot, access-point (AP), scan, and traffic
        actions;
  \item supervised heterogeneous execution through a network management
        system (NMS) coordinating mobility, wireless operations, traffic,
        reporting, safety, and failure handling; and
  \item structured spatial--temporal physical experience that can support OWM
        construction, validation, scenario generation, and preventive network
        management.
\end{enumerate}

The remainder of the paper first develops the OWM viewpoint and explains the
role of \autolab{} within it.  It then presents the common architecture and
end-to-end experiment lifecycle.  Part~I describes the platform from the
Internet researcher's viewpoint, including live observation, historical
reconstruction, experiment authoring, and submission.  Part~II follows the
same experiment beyond the validated contract into NMS registration,
scheduling, physical execution, runtime supervision, measurement, and record
preservation.  An end-to-end trace then demonstrates automatic correction of a
substantial robot movement error, after which the platform is positioned
relative to world models, Digital Twins, shared wireless testbeds, and robotic
measurement systems.

\section{Operational World Models for Wireless Networks}
\label{sec:owm}

\subsection{From World Models to Operational World Models}

A world model forms an internal representation of an environment, predicts how
that environment may evolve, and uses the predicted evolution to support
evaluation or planning.  Recent interest in world models has often been
motivated by visually observed environments, embodied agents, and the prospect
of learning physical dynamics from sensory sequences.  However, neither the
state of a world nor its useful dynamics need to be expressed in the form in
which a human perceives them.  A machine may observe important structures that
are invisible to human sensation, and it may have no operational use for many
details that are visually prominent.

Wireless networks make this distinction particularly clear.  Received-power
fields, channel responses, contention relationships, packet-arrival processes,
service margins, and configuration histories all describe the operating world
of a wireless system.  Most of them cannot be sensed directly by a human, yet
they determine whether traffic can be served and whether a network is
approaching an unstable or unacceptable regime.  Conversely, many visible
details of a room may be unnecessary if they do not materially affect radio
propagation, traffic, or the network decisions under consideration.

We use the term \emph{Operational World Model} to make this machine-centric and
purpose-driven viewpoint explicit.  Our working definition is:

\begin{quote}
An Operational World Model is a machine-centric, task-sufficient
representation of an operational environment that captures relevant spatial
and temporal structures, generates plausible future conditions, evaluates
their operational consequences, and supports preventive actions through
continuous interaction with the physical system.
\end{quote}

This definition does not exclude cameras, floor plans, human observations, or
other familiar sensory inputs.  They can be incorporated whenever they improve
an operational decision.  The definition instead rejects the assumption that
human perception determines what qualifies as a representation of the world.
The operational objective determines which observations and latent structures
the model should retain.

\subsection{Task-Sufficient Representation}

An OWM does not seek a complete digital copy of the physical environment.
Attempting to retain every observable detail would make learning, prediction,
and decision making unnecessarily difficult.  Instead, the model should retain
the information needed for the task while ignoring distinctions that do not
change the relevant future outcomes or actions.

The OWM also does not begin as a blank model when a particular experiment or
deployment starts at $t=0$.  Before this starting time, it may already have
accumulated extensive knowledge of how to interact with the class of worlds in
which it operates.  This knowledge includes both content learned from previous
experience and mechanisms for efficiently interpreting and learning from new
experience.  It plays a role similar to the linguistic and cultural knowledge
that a language model possesses before beginning a conversation with a
particular person.  The observations collected during that conversation refine
the model's understanding of the present context, but they are only a small
portion of its total knowledge.

We therefore distinguish \emph{long-term operational knowledge} from the
\emph{episode history} accumulated after the current interaction begins.  Let
$K_t$ denote long-term knowledge available at time $t$.  It may contain learned
models of spatial and temporal structure, physical and operational constraints,
experience transferred from earlier sites or experiments, and learned rules or
inductive biases governing how new evidence should be absorbed.  Importantly,
$K_0$ is generally rich rather than empty.  Let $h_{0:t}$ denote the much
smaller history of observations, actions, and consequences accumulated in the
current episode, and let $O_t$ denote its most recent observation.  A compact
operational state can then be written as
\begin{equation}
  S_t = f_\theta(K_t,h_{0:t},O_t),
  \label{eq:state}
\end{equation}
where $f_\theta$ uses prior operational knowledge to interpret the current
episode rather than attempting to relearn the world from $h_{0:t}$ alone.

The distinction does not make long-term knowledge immutable.  Experience from
the current episode may later be consolidated into it through a slower update
mechanism,
\begin{equation}
  K_{t+1} = \mathcal{C}_\phi(K_t,h_{0:t},O_t),
  \label{eq:consolidation}
\end{equation}
where $\mathcal{C}_\phi$ represents the learned or designed process by which
new experience modifies long-term memory.  Thus, what is built into the OWM is
not only previously accumulated knowledge, but also a structured way to learn
the world efficiently.  The OWM is not treated as an unstructured predictor
that must learn every relevant relationship from scratch at the start of each
deployment.

Let $A_t$ denote a candidate action and $Y_{t+1:t+T}$ the outcomes of interest
over a future horizon $T$.  The desired $S_t$ is not the largest possible
reconstruction of reality.  It is a state that remains sufficiently
informative for predicting and evaluating the consequences of candidate
actions:
\begin{equation}
  p(Y_{t+1:t+T}\mid K_t,h_{0:t},O_t,A_t)
  \approx
  p_\theta(Y_{t+1:t+T}\mid S_t,A_t).
  \label{eq:sufficiency}
\end{equation}
Equation~\eqref{eq:sufficiency} expresses the paper's principle of
\emph{selective sufficiency}: details irrelevant to the operational objective
may be omitted, even if they are readily sensed or rendered.

In an indoor wireless network, the state may combine several complementary
forms of structure.  A spatial component can be learned from radio maps or
received-power signatures across APs, bands, and antenna configurations.  Such
signatures encode the consequences of geometry, materials, and propagation
without requiring the OWM to reconstruct every physical cause explicitly.  A
temporal component can describe traffic demand at multiple time scales,
associations, contention, and recurring or evolving operating patterns.  The
state may additionally include QoS requirements, AP and station status,
network configuration, and the history of earlier interventions.

The earlier predictive and preventive Digital Twin framework
in~\cite{chen2026predictive} developed this wireless-specific view in detail.
It represented physical space through received-power signatures and statistical
spatial structure, represented traffic through temporally evolving demand, and
used the combined spatial--temporal state to produce physically plausible
future scenarios.  Under the terminology adopted here, those mechanisms are
components of the OWM state and dynamics rather than attempts to reproduce a
complete Physical Twin.

\subsection{Prediction, Evaluation, and Preventive Action}

The purpose of forming $S_t$ is to reason beyond the currently observed
network.  Conditioned on $S_t$ and a possible action sequence, the OWM produces
a distribution of plausible future scenarios,
\begin{equation}
  \widetilde{S}_{t+1:t+T}
  \sim
  p_\theta\!\left(
    S_{t+1:t+T}\mid S_t,A_{t:t+T-1}
  \right).
  \label{eq:future}
\end{equation}
The word \emph{plausible} is essential.  Infinitely many hypothetical scenarios
can be constructed that would break a wireless network, but scenarios that
violate physical structure or have negligible probability are not useful
targets for preventive management.  The role of scenario generation is
therefore not unrestricted imagination; it is to explore future conditions
consistent with learned spatial--temporal structure and current evidence.

A risk or service-quality functional $R$ evaluates these futures under the
candidate actions.  In abstract form, preventive planning selects
\begin{equation}
 A^*_{t:t+T-1}
 = \arg\min_{A_{t:t+T-1}}
 \mathbb{E}\!\left[
 R\!\left(\widetilde{S}_{t+1:t+T},A_{t:t+T-1}\right)
 \mid S_t
 \right].
 \label{eq:planning}
\end{equation}
The objective in~\eqref{eq:planning} need not seek a fragile global optimum.
For operational management, a more useful goal may be to detect that a future
scenario approaches a feasibility boundary and then restore adequate service
margin before degradation occurs.

The framework in~\cite{chen2026predictive} instantiated this principle using
two efficiently computable feasibility screens: a Shannon-capacity-based bound
and a CSMA/CA-latency-based bound.  Importance sampling was used to concentrate
evaluation on risky generated scenarios, and preventive reconfiguration was
directed toward restoring performance margins through choices such as AP--STA
association and backhaul configuration.  These analytical and algorithmic
details remain supporting theoretical work; the present paper focuses on the
physical experimentation infrastructure required to ground, exercise, and
extend such an OWM.

The selected intervention is executed in the physical network, and the
resulting measurements are appended to $H_{t+1}$.  Representation, future
generation, operational evaluation, planning, intervention, and feedback
therefore form a continuing experience loop.  Prediction alone is insufficient:
an operational model must be able to compare predicted consequences with the
outcomes of controlled actions in the real system.

\subsection{Relationship to Digital Twins}

Digital Twins and OWMs are related rather than mutually exclusive.  Digital
Twin work commonly emphasizes correspondence, replication, and synchronization
between a physical system and a digital representation.  These capabilities
are useful to an OWM, but they do not by themselves express its central
intention.

An OWM is organized around the future operating consequences of the present
state.  It asks what information from history and current observation is needed
to generate relevant future scenarios, how the scenarios should be evaluated,
which preventive intervention should be taken, and what can be learned from the
observed outcome.  Physical replication is therefore one possible source of
state and simulation capability, not the defining endpoint.

This terminology clarifies rather than changes the direction of the work in
\cite{chen2026predictive}.  That work already looked into the future using
historical experience and current scenario state, evaluated predicted futures,
and sought to intervene before performance degradation.  The OWM name makes
this original predictive and preventive intention explicit and places it in a
broader machine-centric framework.

\subsection{Role of AutoLab}

An OWM cannot be grounded solely by an offline dataset or by simulations whose
assumptions never confront a changing physical network.  It requires a
mechanism for creating controlled experience: observing a real environment,
altering selected operating conditions, recording both the intervention and
its consequences, and making the experiment repeatable.

\autolab{} provides this mechanism.  Through an Internet-facing research
interface and a physically deployed NMS, it supports time-ordered experiments
involving mobile robots, wireless scans, AP operations, and traffic sessions.
It validates and schedules the requested actions, supervises their execution,
and records measurements, network state, command outcomes, mobility traces, and
diagnostics.  The resulting data are structured records of how a physical
wireless world responded to controlled actions under known scenario conditions.

Within the complete OWM loop, \autolab{} therefore serves four closely related
functions:
\begin{enumerate}
  \item \emph{observation}: collecting machine-native measurements and network
        state from the physical environment;
  \item \emph{intervention}: imposing reproducible changes in mobility,
        traffic, scanning, and network configuration;
  \item \emph{grounding}: comparing model-generated conditions and predicted
        consequences with physical outcomes; and
  \item \emph{experience accumulation}: preserving synchronized states,
        actions, and outcomes for subsequent learning and evaluation.
\end{enumerate}

The distinction between present capability and intended integration is
important.  \autolab{} already implements substantial observation,
intervention, execution, and feedback infrastructure.  Generative scenario
production and predictive and preventive network-management algorithms are
supporting work under continuing development.  Accordingly, this paper does
not claim that a complete autonomous OWM is already deployed.  It presents
\autolab{} as a major physical-system step in constructing and validating that
larger system.

\section{AutoLab Architecture and End-to-End Experiment Lifecycle}
\label{sec:architecture}

Section~\ref{sec:owm} described the knowledge, prediction, evaluation, action,
and feedback functions of an OWM.  \autolab{} supplies the system structure
through which these functions can interact with a real wireless environment.
Its purpose is not limited to remote device control.  It connects an Internet
researcher or a future OWM decision process to a physical test site through a
repeatable experiment lifecycle: the intended scenario is specified, checked,
admitted, executed, observed, and preserved as reusable experience.

This section presents the common architecture of that lifecycle.  It provides
the seam between the OWM motivation and the two implementation views developed
later in the paper.  Part~I follows the lifecycle from the viewpoint of an
Internet researcher, whereas Part~II follows the same experiment from the NMS
and physical-site viewpoint.

\subsection{Functional Architecture}

\autolab{} is organized into six functional layers, summarized in
Table~\ref{tab:layers}.  These layers describe responsibilities rather than a
requirement that every function execute in a separate computer or software
process.  The first five layers substantially describe the present platform.
The sixth places the platform within the continuing OWM program and contains a
combination of earlier theoretical results, algorithms under development, and
future system integration.

\begin{table}[t]
\centering
\caption{Functional layers of AutoLab within the OWM architecture.}
\label{tab:layers}
\begin{tabular}{p{0.23\textwidth}p{0.67\textwidth}}
\toprule
Layer & Primary responsibility \\
\midrule
Researcher interaction & Observe the site, inspect history, construct experiments, and retrieve results. \\
Experiment specification & Describe initial conditions and time-ordered robot, AP, scan, and traffic actions. \\
Validation and registration & Apply common and site-specific rules and admit valid experiments. \\
Physical orchestration & Schedule commands and supervise devices, mobility, traffic, and safety. \\
Observation and history & Collect measurements, NMS state, command outcomes, traces, and diagnostics. \\
OWM learning and decision & Learn operational state, generate plausible futures, evaluate risk, and plan preventive action. \\
\bottomrule
\end{tabular}
\end{table}

The \emph{researcher-interaction layer} provides the human-facing view of the
physical site.  It allows a researcher to inspect current conditions and
recorded history, prepare an experiment, and retrieve the resulting records.
This layer hides internal device details that are unnecessary to experiment
design while exposing the operational state needed to understand what is
happening at the site.

The \emph{experiment-specification layer} converts an experimental intention
into explicit initial conditions and time-ordered actions.  Rather than
teleoperating motors, radios, or individual software processes, a researcher
states what should happen: for example, where a robot should move, when a scan
should occur, which bounded AP operation should be applied, or when a traffic
session should run.  Timing relationships among these heterogeneous actions
define the scenario to be reproduced.

The \emph{validation-and-registration layer} determines whether the requested
scenario is meaningful and admissible at a particular site.  It applies both
rules common to \autolab{} and restrictions arising from the site's geometry,
devices, and operating policy.  Registration associates an accepted
specification with a scheduled experiment instance.

The \emph{physical-orchestration layer} is centered on the NMS.  It translates
the admitted experiment into scheduled commands, delivers them to the
appropriate robot or AP agents, coordinates traffic and scanning operations,
and supervises physical execution.  Destination-level mobility illustrates the
role of this layer: the experiment specifies a desired pose, whereas the
physical system performs location estimation, turning, forward movement,
evaluation, correction, and controlled stopping.

The \emph{observation-and-history layer} collects the state and consequences
of execution.  Its records include measurements, device and NMS states,
command outcomes, traffic results, mobility traces, and diagnostic information.
Because the requested action, its timing, the surrounding state, and the
measured outcome are retained together, the result is structured operational
experience rather than an isolated collection of measurement files.

The \emph{OWM-learning-and-decision layer} uses such experience to improve the
long-term knowledge $K_t$ introduced in Section~\ref{sec:owm}, infer current
operational state, generate plausible future scenarios, evaluate their risks,
and propose preventive actions.  The distinction in implementation status is
important: \autolab{} presently provides substantial infrastructure in the
first five layers, while full integration of the sixth layer remains continuing
work.

\subsection{The Validated Experiment Contract}

The public-facing and physical-site portions of the system meet through a
\emph{validated experiment contract}.  The public authoring tool produces two
explicit tables: one describes the time-ordered command sequence, and the other
describes the initial poses required by the experiment.  These tables are
exported as \texttt{CommandSheet.csv} and \texttt{InitialPoses.csv}.  Together
with the selected site and scheduled starting time, they define what the NMS is
being asked to execute.

The exchange format alone is not sufficient to protect a physical laboratory.
A specification may be syntactically well formed while requesting an unknown
device, an invalid argument, conflicting timing, an unsafe movement, or an
action prohibited by the selected site.  \autolab{} therefore treats
validation as part of the contract rather than as a convenience of the user
interface.

The same authoritative Python \emph{CommonCheckers} logic is invoked at two
boundaries.  It first runs as a public preflight while the researcher is
preparing the experiment.  It then runs again when the lab operator registers
the exported files with the NMS.  The workbook macros provide pull-down lists,
construct argument JSON, draw planned movements, and export data, but they do
not decide whether an experiment passes or fails.  Consequently, bypassing a
GUI aid or manually modifying an exported file does not bypass the rules used
for physical admission.

This repeated use of one implementation establishes a stable semantic boundary
between the two parts of the system.  A successful public preflight gives the
researcher early and specific feedback, while validation at registration
protects the physical site using the same interpretation of the rules.

\subsection{End-to-End Experiment Lifecycle}

An \autolab{} experiment proceeds through the following stages.

\begin{enumerate}
  \item \emph{Observe and formulate.}  The researcher inspects the site and
        available historical information, then defines the question or
        scenario to be tested.
  \item \emph{Specify.}  Initial robot poses and time-ordered robot, scan, AP,
        and traffic commands are entered through the public authoring tool.
  \item \emph{Preflight.}  Common and site-specific rules are evaluated before
        submission, allowing errors to be corrected without occupying the
        physical site.
  \item \emph{Register and schedule.}  The exported contract is checked again,
        associated with an experiment identity and absolute start time, and
        admitted to the NMS schedule.
  \item \emph{Execute and supervise.}  Relative command offsets determine the
        physical sequence.  The NMS coordinates device agents, observes their
        progress, and applies the runtime procedures needed to complete or
        safely terminate each action.
  \item \emph{Report and preserve.}  Measurements, states, outcomes, traces,
        and diagnostics are returned to the supporting services and retained
        as the history of the experiment.
  \item \emph{Inspect and learn.}  Researchers retrieve or reconstruct the
        experiment, while OWM algorithms may use the structured experience for
        model learning, prediction assessment, or later action design.
  \item \emph{Endure unknown futures.}  The accumulated knowledge is tested in
        scenarios that were not revealed to the model in advance.  Independent
        researchers or scenario-generating systems define the conditions and
        disturbances, while competing models must observe, adapt, and act to
        maintain acceptable operation.  The objective is not to memorize how
        to survive one fixed scenario, but to acquire transferable capability
        for enduring previously unseen conditions at the same site and,
        ultimately, at different sites.
\end{enumerate}

\begin{figure}[H]
\centering
\includegraphics[width=\textwidth,height=0.86\textheight,keepaspectratio]{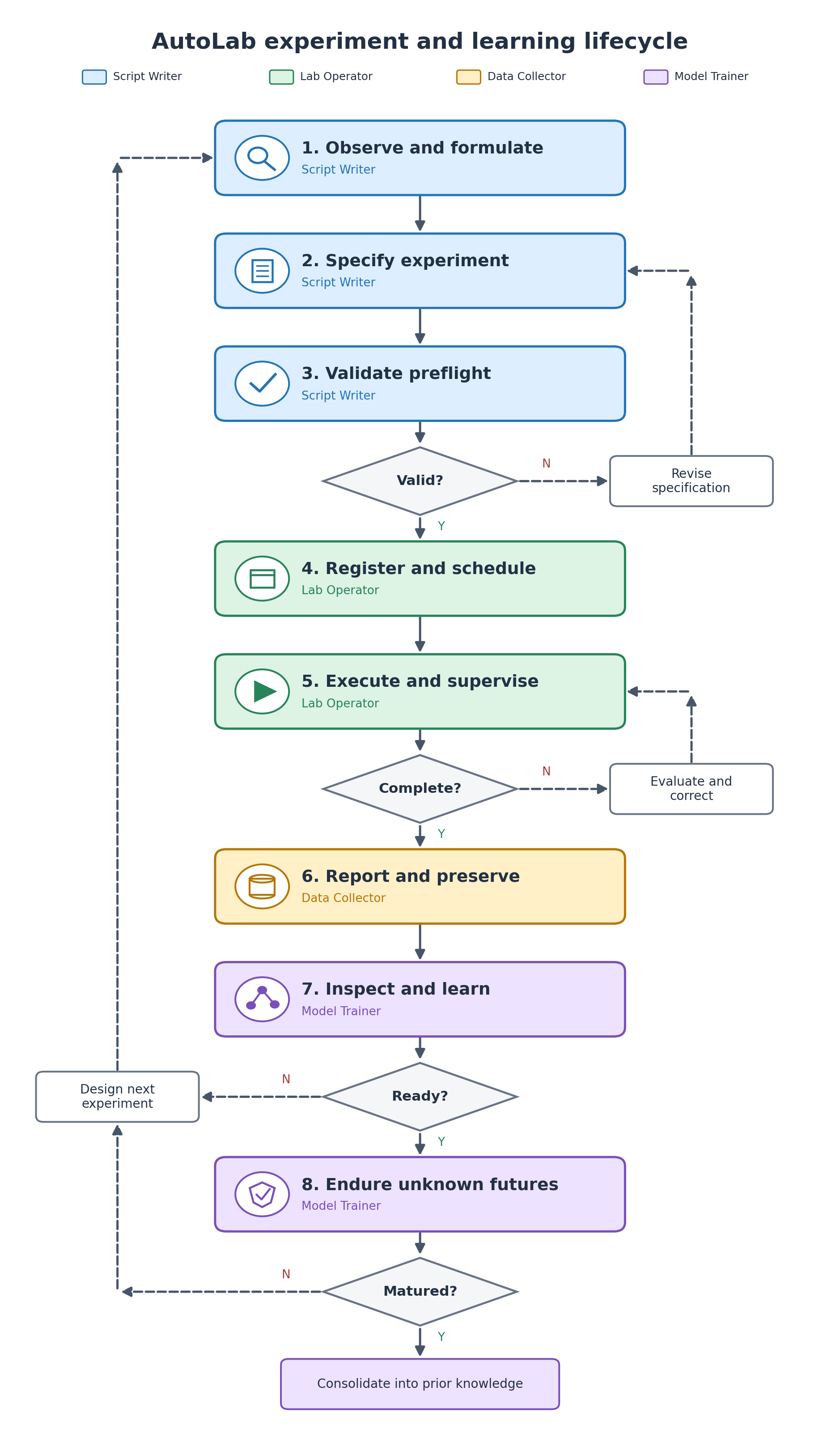}
\caption{Eight-stage \autolab{} experiment and learning lifecycle.  Short
dashed look-back paths denote local iteration until a satisfactory result is
obtained.  The outer return path from Stage 8 indicates that experience gained
while confronting unknown futures becomes prior knowledge for later
interaction.}
\label{fig:lifecycle}
\end{figure}

Figure~\ref{fig:lifecycle} summarizes the eight stages.  The forward arrows
show the nominal progression of an experiment, while the smaller dashed
look-back paths show iterations that may continue until their local objective
is satisfied.  A specification may be revised until it passes validation,
physical actions may be repeatedly evaluated and corrected under runtime
supervision, and inspection of the results may motivate another controlled
experiment.  Stages 1--7 thereby form the cycle through which controlled
experience is created and absorbed.  Stage 8 is the larger objective of that
learning cycle: using accumulated operational knowledge to function under a
future that is not specified to the model in advance.

The direction from specification to physical action is therefore paired with a
return path from physical outcome to observation and long-term knowledge.  The
experiment lifecycle can be summarized as
\begin{equation}
\begin{split}
\text{intention} &\rightarrow \text{validated specification}
\rightarrow \text{physical intervention}\\
&\rightarrow \text{measured outcome}
\rightarrow \text{operational experience}.
\end{split}
\label{eq:lifecycle}
\end{equation}

\subsection{AutoLab as an OWM Experience Interface}

From the viewpoint of Section~\ref{sec:owm}, an experiment is an intentional
interaction with the world.  The initial conditions and command sequence define
the intervention, the physical site supplies dynamics not fully captured by an
offline model, and the observation-and-history layer records the response.
Repeated experiments can therefore test whether a predicted scenario is
physically plausible, assess whether a proposed action has the intended
consequence, and add new evidence to long-term operational knowledge.

The ultimate purpose is broader than adapting passively to one known test site
or learning a policy for one prescribed scenario.  \autolab{} separates the
party that defines an experimental future from the model that must function
within it.  Researchers, automated generators, or other models can construct
previously unseen combinations of traffic, mobility, radio conditions, and
network interventions.  A model under evaluation receives the observations
available during execution, but not advance knowledge of the complete future
scenario.  Its task is to use prior knowledge, absorb new evidence quickly, and
maintain acceptable operation as the scenario unfolds.

This separation changes the meaning of experimental success.  Success is not
merely reproducing a known script or learning a response tailored to it.  It is
the development of reusable operational competence: the ability to confront
unknown hardship, first within a familiar site and later in environments whose
geometry, devices, or operating patterns differ from those previously seen.
Such evaluation requires systematic scenario administration, observation,
intervention, outcome recording, and bookkeeping across models and experiments.
These are precisely the forms of physical infrastructure and persistent
experience supplied by \autolab{}.

At present, a human researcher commonly supplies the experimental intention and
interprets the returned result.  The same architecture also provides a future
interface for OWM algorithms: a generated scenario or preventive action can be
translated into an admissible experiment, grounded through supervised physical
execution, and compared with its predicted outcome.  This future integration
does not require weakening the validation boundary.  Model-generated
experiments must satisfy the same explicit contract and physical-site rules as
human-authored experiments.

This contract makes the two views in the remainder of the paper parts of one
closed lifecycle rather than separate subsystems.  Part~I follows an experiment
from the researcher's Internet-facing viewpoint.  Part~II follows the accepted
experiment through NMS registration, physical execution, and reporting.

\begin{figure}[H]
  \centering
  \includegraphics[width=\textwidth,height=0.84\textheight,keepaspectratio]{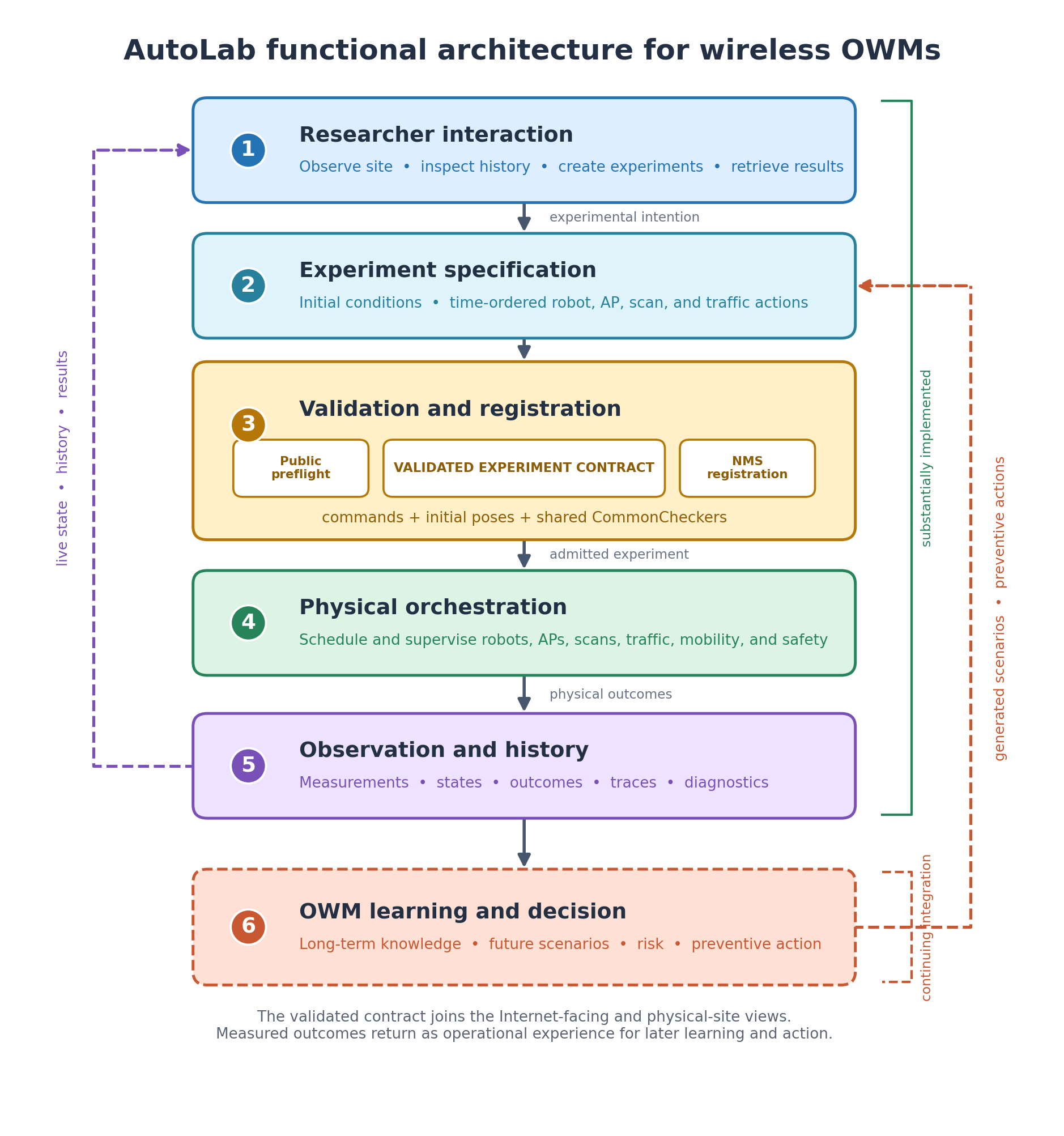}
  \caption{Functional architecture of \autolab{} within the wireless OWM
  framework.  The validated experiment contract joins public experiment
  specification to NMS-controlled physical execution.  Physical outcomes
  return through observation and history as operational experience, while
  learned scenarios and preventive actions can initiate later experiments.
  Solid layer borders denote the substantially implemented \autolab{} platform;
  the dashed sixth layer denotes continuing OWM integration.}
  \label{fig:architecture}
\end{figure}

\part{Internet Researcher's View}

\section{Live Observation of the Physical Test Site}
\label{sec:live}

The first Internet-facing role of \autolab{} is to make a running physical
experiment intelligible without requiring the observer to log into the NMS or
individual devices.  The deployed web interface therefore presents the test
site as a continuously updated operational scene rather than as a collection
of independent status pages.  A site identifier reported by the NMS selects
the corresponding site map.  For the present deployment,
the identifier \texttt{DemoRoom} selects the DemoRoom map.  Robot locations
reported in meters and fixed AP locations are rendered in the same physical
coordinate system, allowing the Internet view to preserve their spatial
relationship as the experiment evolves.

The Internet-facing laboratory interface is publicly presented at
\url{https://www.6g-private.com/lab}.  Short demonstration and tutorial videos
for the public interface are maintained at
\url{https://www.youtube.com/@Hands-OnDigitalTwins}.

\begin{figure}[t]
\centering
\IfFileExists{fig3_live_interface.png}{%
  \includegraphics[width=0.98\textwidth]{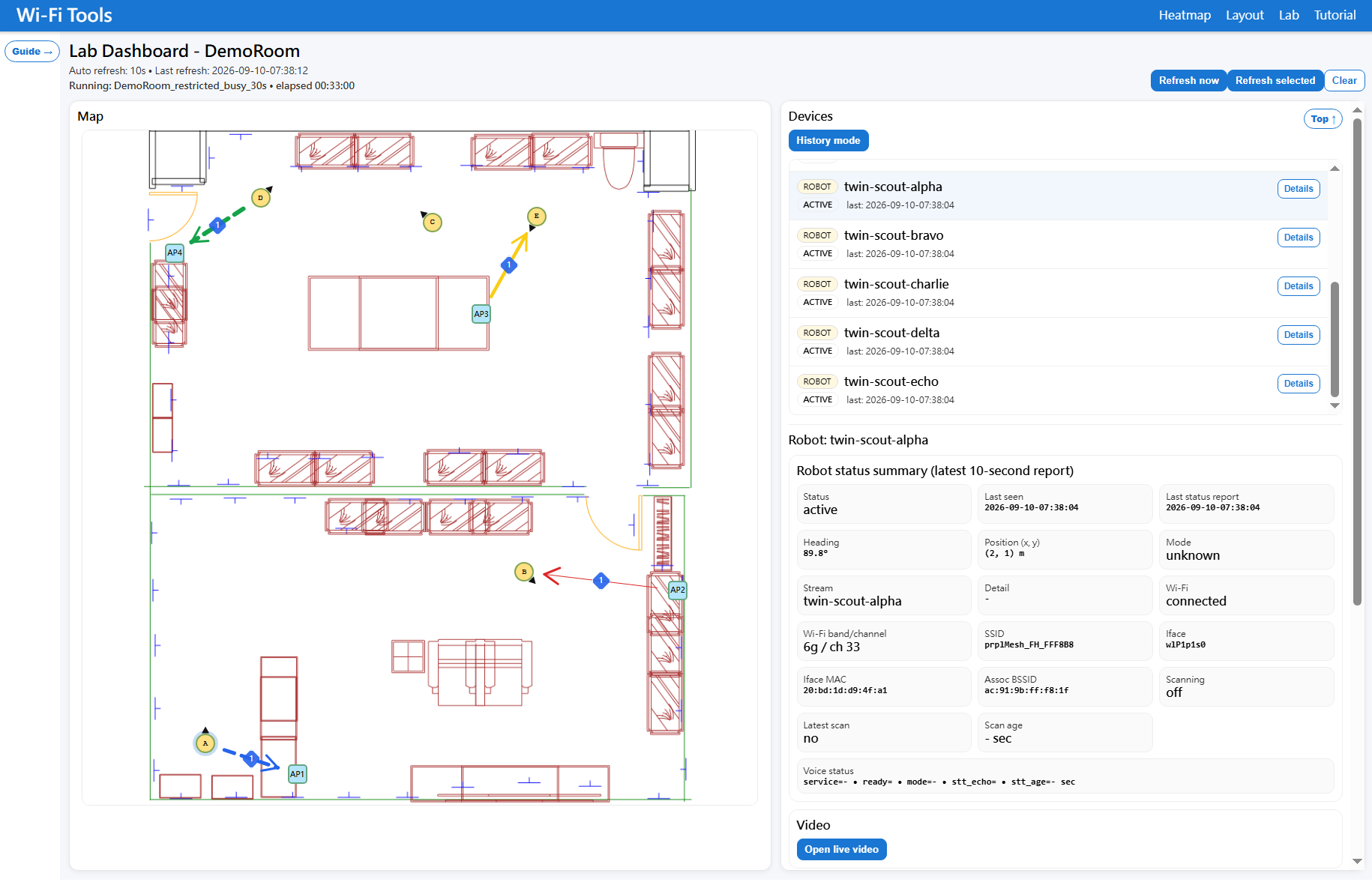}%
}{%
  \fbox{\parbox[c][0.28\textheight][c]{0.92\textwidth}{\centering
  Placeholder for \texttt{fig3\_live\_interface.png}: full LIVE-page screenshot
  showing the Lab Observer Guide, DemoRoom map, robots/APs and active traffic,
  and the right-hand status panel.}}%
}
\caption{Internet-facing LIVE view of an AutoLab experiment.  The interface
combines the contextual Lab Observer Guide, a spatial map of the physical test
site with robots, APs, associations and active traffic, and selectable device
and session details in the right-hand panel.}
\label{fig:live-interface}
\end{figure}

Figure~\ref{fig:live-interface} shows the Internet-facing LIVE interface used to
observe an active DemoRoom experiment.

The center panel provides the primary spatial view.  Robots and APs are placed
on the site map using the locations contained in the current site state.  The
visual treatment of a device communicates coarse status before the user opens
its detailed record.  Robot labels distinguish disconnected devices from
currently usable devices and emphasize an active Wi-Fi association; AP labels
similarly distinguish offline from available APs.  These visual cues are not
substitutes for the underlying status fields.  They are a compact first layer
for deciding where a researcher should look next.  Selecting a device exposes
its detailed state in the right-hand panel while leaving the spatial context
visible.

For a robot, the detail panel presents the information needed to relate
physical movement to wireless operation.  This includes the reported pose,
robot status, Wi-Fi interface state, SSID and associated BSSID when available,
operating frequency or band, scanning state, and other status fields carried
by the periodic report.  For an AP, the same panel exposes its fixed location,
overall state, current associations, and per-interface information.  The AP
interface representation has been extended beyond band and channel to retain
fields such as interface name, BSSID, SSID, center frequency, and channel
width.  Detailed machine-readable information is preserved behind the compact
human-oriented summary so that an experienced researcher can inspect the
reported state rather than relying only on the graphical encoding.

Active traffic is superimposed directly on the same map.  A traffic session is
represented by an arrow between the robot and AP involved in that session.
Direction indicates the direction of the generated flow.  The line color is
derived from the Wi-Fi access category, the solid or dashed line style
distinguishes TCP from UDP, and line width represents the configured
traffic rate when that rate is available; otherwise, the display uses a
defined access-category-based fallback width.  The arrow head and line share the access-category
color but the traffic-rate encoding is applied to the line rather than being
allowed to enlarge the arrow head.  A numbered marker enclosed by a diamond is placed at the center of the traffic
arrow.  The researcher selects this marker to expose the session records for
that robot--AP traffic relationship, including session identifier, protocol,
access category, direction, duration, configured-rate fields, status, and the
underlying event information.  When multiple sessions are simultaneously active
between the same robot and AP, the number identifies how many active session
records are represented by that map relationship; selecting the diamond exposes
the individual records.  The resulting view allows spatial movement,
association, and application traffic to be inspected in one scene instead of
being reconstructed manually from separate logs.

\begin{table}[t]
\centering
\caption{Implemented information available from the Internet-facing LIVE view.}
\label{tab:liveview}
\begin{tabular}{p{0.20\textwidth}p{0.34\textwidth}p{0.36\textwidth}}
\toprule
Object & Visible or selectable information & Operational use \\
\midrule
Site/NMS & Site identity, current experiment state, NMS status & Establish which physical site and experiment are being observed. \\
Robot & Position, connectivity, Wi-Fi state, association, scan/status details & Relate mobility and device state to network behavior. \\
AP & Fixed position, state, interfaces, SSID/BSSID, band/channel/frequency/width, associations & Inspect infrastructure state and the AP-side view of connectivity. \\
Traffic session & Endpoints, direction, protocol, access category, duration, rate-related fields, status & Relate offered traffic and QoS treatment to the physical topology. \\
Robot camera & Live physical scene from the selected robot & Cross-check the digital site view against the physical environment. \\
\bottomrule
\end{tabular}
\end{table}

The map is supplemented by a direct physical view from robot-mounted cameras.
An Internet observer can select an available camera stream and see the test
site from the robot's perspective while the experiment continues.  This is
particularly useful when the structured map indicates an unexpected condition:
the researcher can compare reported pose and network state with the scene that
the robot is physically observing.  Camera imagery is therefore used as a
complementary observation channel rather than as the coordinate system of the
experiment.  Different Internet users may simultaneously request camera views
from different robots.  To bound the required public-network bandwidth, video
is broadcast only on demand and each request keeps the broadcast available for
one minute rather than enabling an indefinite public stream.

\begin{figure}[p]
\centering
\begin{subfigure}[t]{\textwidth}
  \centering
  \includegraphics[width=\linewidth]{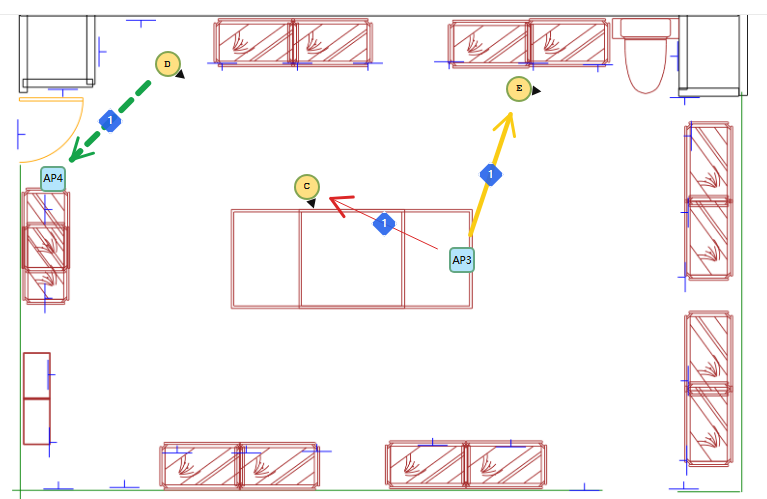}
  \caption{LIVE map view.}
  \label{fig:robot-camera-map}
\end{subfigure}

\vspace{0.8em}
\begin{subfigure}[t]{\textwidth}
  \centering
  \includegraphics[width=\linewidth]{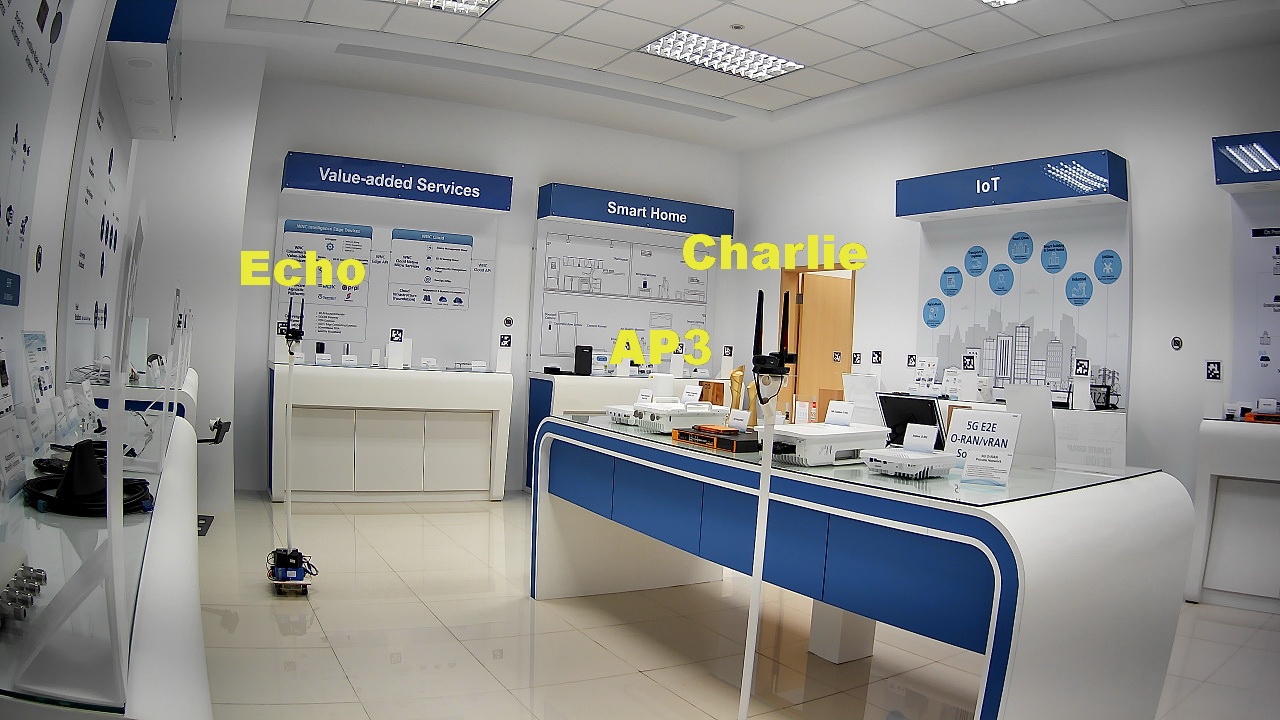}
  \caption{On-demand robot-camera view.}
  \label{fig:robot-camera-view}
\end{subfigure}
\caption{Spatial and physical views from the same running experiment.  The
LIVE map in (a) provides site-relative robot/AP locations, headings, and network
context, while the on-demand camera view in (b) provides direct physical
context.  Named objects visible in the camera image, including AP3 and the
nearby robots Charlie and Echo, can be matched to their corresponding objects
on the site map, providing a direct cross-check between reported spatial state
and the physical scene.}
\label{fig:robot-camera}
\end{figure}

The relationship between the map location and heading and the robot's physical view is
illustrated in Fig.~\ref{fig:robot-camera}.  In this example, objects visible
from the robot camera---including AP3 and the nearby robots Charlie and
Echo---also appear on the LIVE map, so the two views can be matched using
physical landmarks and device identities.

The interface is organized for both first-time observers and technical users.
The left panel contains a contextual \emph{Lab Observer Guide} whose text is
updated after important user actions and suggests what can be inspected next.
It can be collapsed when more map area is desired.  The center map and the
right information panel scroll independently so that detailed records can be
read without losing the spatial view; selecting a map object brings its detail
block into view, and a floating control returns the information panel to its
top.  These mechanisms do not alter experiment execution.  They are
observation aids intended to make a running physical experiment understandable
from an ordinary web browser.  The public tutorials linked above show these
interactions using the deployed interface rather than a separate mock-up.

The LIVE page is driven by periodically refreshed site state rather than by
browser-side inference about the physical system.  In the current
implementation, the web server consumes the latest NMS status snapshot and
constructs a dashboard representation containing the site/NMS identity,
experiment state, device records, AP--robot associations, and active traffic
sessions.  The browser polls this representation and renders the newest state.
Thus, the public visualization and the NMS-side execution remain separated:
the website observes and presents the experiment, while physical control and
runtime supervision remain at the test site as described in Part~II.

\section{Historical State Reconstruction and Replay}
\label{sec:history}

A live view is insufficient when an important transition occurred before the
researcher looked at the page, when several events must be correlated in time,
or when collaborators want to inspect an experiment after its execution.
\autolab{} therefore complements the current dashboard with a history mode.
Throughout this paper, \emph{replay} means reconstruction and interactive
inspection of recorded historical state.  It does not mean that the physical
robots and APs execute the experiment a second time.

The web-server history service records and digests the periodic NMS state used
by the dashboard.  A historical state can consequently contain the same kinds
of objects used by the LIVE view---site identity, experiment state, robot and
AP status, locations, associations, and traffic-session state---at an earlier
observation time.  History is therefore structured state rather than a video
recording of the webpage.  The distinction is important: when the user moves
to an earlier time, the reconstructed robot, AP, and traffic objects remain
selectable and their recorded fields can still be inspected.

The HISTORY interface places time-oriented controls above the object details.
A researcher first requests the historical interval of interest and then uses
the playback controls and event log to navigate it.  Playback advances the
reconstructed site state through recorded time; the user can pause and inspect
a particular moment rather than allowing the timeline to continue.  The event
log provides a complementary event-oriented route into the same history, which
is useful when the researcher knows what transition is interesting but not its
exact timestamp.  Device and traffic details appear below these controls, so
temporal navigation remains accessible while individual objects are examined.

\begin{figure}[t]
\centering
\IfFileExists{fig5_history_interface.png}{%
  \includegraphics[width=0.98\textwidth]{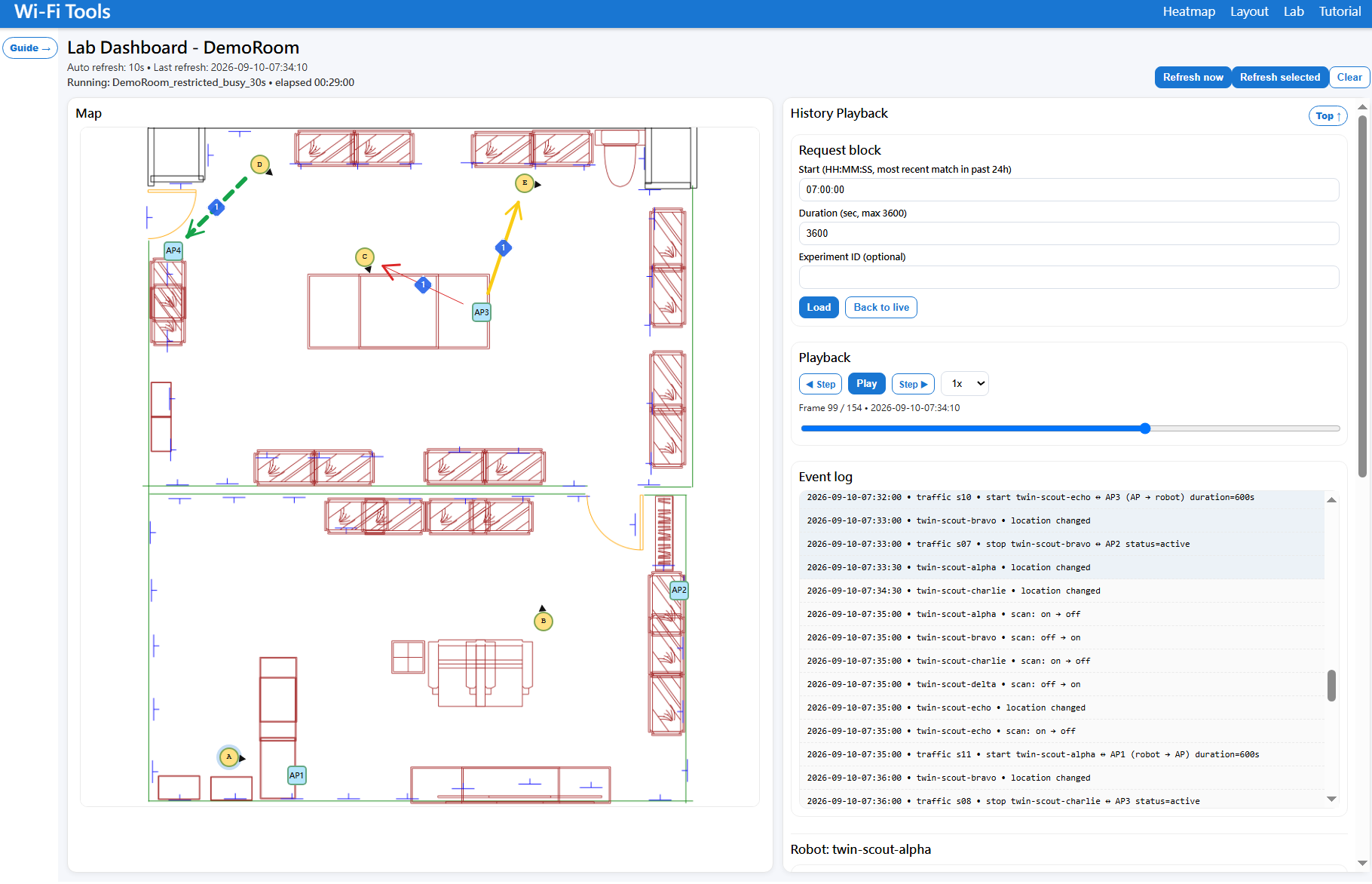}%
}{%
  \fbox{\parbox[c][0.28\textheight][c]{0.92\textwidth}{\centering
  Placeholder for \texttt{fig5\_history\_interface.png}: HISTORY-page
  screenshot showing Request, Playback, Event Log, the reconstructed map, and
  a selected robot or traffic-session record.}}%
}
\caption{Historical state reconstruction and replay in AutoLab.  The HISTORY
view combines interval selection, playback controls and event records with the
same structured spatial, device and traffic information used by the LIVE view.}
\label{fig:history-interface}
\end{figure}

Figure~\ref{fig:history-interface} shows the HISTORY interface used for this
interactive reconstruction.

This representation supports a form of temporal debugging that would be
awkward with independent text logs.  For example, a researcher can navigate to
the interval in which a robot moved, inspect the robot's recorded wireless
state, examine which AP was associated with it, and inspect the traffic
sessions active at that time.  The map supplies the spatial context while the
right panel supplies the recorded device and session fields.  Because these
views are derived from a common historical state, the researcher does not have
to align several logs manually merely to establish what the system believed at
a given instant.

Virtual scripted scenarios are independent of the history mechanism.  When no
physical experiment is active or scheduled to begin soon, a periodic scripted
scenario can instead be presented through the same LIVE interface as a regular
introduction to the public website.  From the dashboard's viewpoint it produces
the same form of site state used for live presentation, so visitors can explore
the map, device state, and traffic visualization without waiting for a physical
experiment.  The LIVE page identifies this condition explicitly as virtual mode
so that it cannot be mistaken for current physical execution.

Because the history service records the LIVE state stream, a virtual scenario
can also appear later in HISTORY in exactly the same way as a physical
experiment.  HISTORY intentionally does not classify a reconstructed state as
virtual or physical: its role is to reconstruct and expose the state that was
presented at that time.  Thus, replay remains independent of the mechanism that
originally generated the LIVE state.  Claims in this paper about physical
execution are supported by physical logs, screenshots, or demonstrations rather
than by the introductory virtual scenario.

Historical dashboard data are available for 24 hours.  This window is intended
for operational verification rather than permanent archival storage: a remote
researcher who wants to confirm that an experiment proceeded as expected can
return within one day, request the relevant interval, and inspect the recorded
spatial, device, association, traffic, and event state.  Experiment result data
and runtime records follow the experiment-record path described in the
end-to-end lifecycle and in Part~II; the HISTORY page is the short-term
interactive reconstruction interface rather than the permanent experiment-data
repository.

\section{Internet Experiment Authoring and Submission}
\label{sec:authoring}

Observation and history make a physical experiment visible, but an
Internet-accessible experimental platform must also provide a reproducible way
to express what the site should do.  \autolab{} uses a script-template workflow
in which a researcher specifies initial conditions and a time-ordered sequence
of supported operations.  The authoring tool is an Excel macro-enabled template distributed together
with public Python validation code and site-specific assets.  The downloadable
script-template package is provided with the AutoLab authoring tutorial at
\url{https://www.youtube.com/watch?v=3PegjtwTM5o\&t=6s}.  The spreadsheet is an authoring interface; the exported
experiment contract remains the two explicit tables introduced in
Section~\ref{sec:architecture}.

The first table, \texttt{CommandSheet}, contains the ordered experiment
commands.  Each row identifies when an action should occur, the command family
and target device, and the command arguments.  The second table,
\texttt{InitialPoses}, records the initial robot poses and the site/device
information required to establish the starting condition.  The template
constructs the machine-readable \texttt{args\_json} field from the values
entered by the author rather than requiring the researcher to compose JSON
manually.  The resulting tables are exported as \texttt{CommandSheet.csv} and
\texttt{InitialPoses.csv}, which form the public-side representation handed to
registration.

The implemented command vocabulary is deliberately bounded.  Robot mobility
commands include destination movement, location reporting, and controlled
transitions between the inner and outer portions of the DemoRoom site.
Wireless scan commands provide one-shot, start, and stop operations.  AP
commands presently include bounded station-disassociation and transmit-power
operations.  Traffic authoring supports TCP and UDP traffic-session start
commands with explicit parameters such as session identity, duration, access
category, payload/rate settings as applicable, and target robot.  The
vocabulary is extended by adding explicit command semantics and validation
rules rather than by allowing arbitrary remote shell commands.

\begin{table}[t]
\centering
\caption{Implemented public authoring functions and their authority.}
\label{tab:authoring}
\begin{tabular}{p{0.24\textwidth}p{0.40\textwidth}p{0.26\textwidth}}
\toprule
Function & Present role & Determines acceptance? \\
\midrule
Spreadsheet GUI & Command/device dropdowns, parameter entry, defaults, and convenient editing & No \\
\texttt{BuildArgsJson()} & Constructs \texttt{args\_json} from author-entered fields & No \\
Map guidance & Draws planned poses/paths and site context for the selected command & No \\
\texttt{CommonCheckers} & Applies public Python command, argument, device, and site rules & Yes \\
CSV export & Produces \texttt{CommandSheet.csv} and \texttt{InitialPoses.csv} & No \\
Registration handoff & Submits the exported contract to the laboratory-side registration path & Acceptance is rechecked by the same Python validation rules \\
\bottomrule
\end{tabular}
\end{table}

The spreadsheet provides site-aware guidance while the experiment is being
written.  Device dropdowns constrain ordinary selection to appropriate robot
or AP names; parameter cells are expanded according to the selected command;
and the map view can display planned robot poses and paths relative to the
site geometry.  These mechanisms reduce authoring errors and make a script
easier to understand visually.  They are intentionally not the authoritative
safety boundary.  In particular, map drawing and VBA macros do not determine
whether an experiment passes validation.

Authoritative preflight validation is implemented in the public Python
\texttt{CommonCheckers} package.  The script-writer package separates common
validation code from site assets; for the present site, DemoRoom-specific
configuration is selected using the laboratory identity.  The checkers verify
the exported tables against the supported command vocabulary, argument rules,
device/whitelist constraints, and applicable site restrictions.  Invalid
parameters may therefore be representable temporarily in the GUI but are
rejected by Python before the experiment is considered valid.  This separation
keeps authoring convenience from becoming a second, inconsistent source of
validation truth.

The same validation logic is used again at the laboratory boundary.  The
script writer can invoke \texttt{RunCommonCheckers()} locally, which calls the
Python checker and reports PASS/FAIL information before submission.  At
registration, the web/API path accepts the two exported CSV files and invokes
the same \texttt{CommonCheckers} rules rather than trusting the result of a VBA
macro or a validation implementation duplicated in the web server.  The two
checks occur at different trust boundaries but use the same rule source.  A
script that was edited after local checking, targeted at an incompatible site,
or otherwise fails the authoritative rules is therefore not admitted merely
because the spreadsheet previously displayed a successful preflight result.

\begin{figure}[p]
\centering
\begin{subfigure}[t]{0.98\textwidth}
  \centering
  \includegraphics[width=\linewidth]{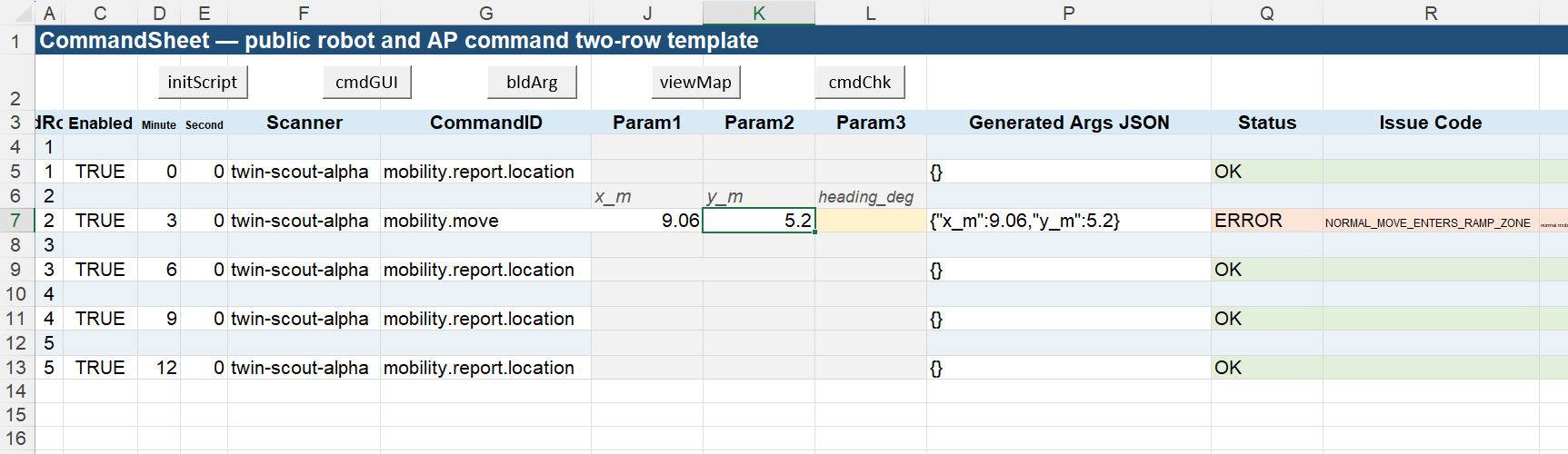}
  \caption{\texttt{CommandSheet}: time-ordered commands, generated arguments,
  and validation status.}
  \label{fig:authoring-command}
\end{subfigure}

\vspace{0.7em}
\begin{subfigure}[t]{0.98\textwidth}
  \centering
  \includegraphics[width=\linewidth]{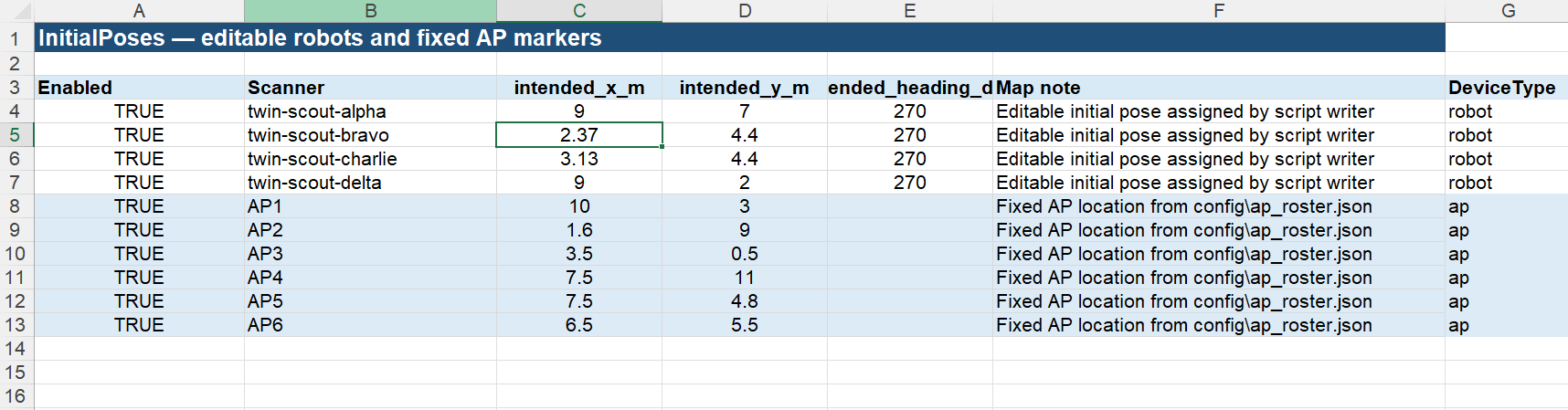}
  \caption{\texttt{InitialPoses}: editable robot starting poses and fixed AP
  positions loaded from site configuration.}
  \label{fig:authoring-init}
\end{subfigure}
\caption{AutoLab script-template authoring interface: (a) the researcher enters
 time-ordered operations in \texttt{CommandSheet}, where arguments and validation
 results are shown, and (b) defines experiment starting conditions in
 \texttt{InitialPoses}.}
\label{fig:authoring-interface}
\end{figure}

\begin{figure}[p]
\centering
\includegraphics[width=0.98\textwidth]{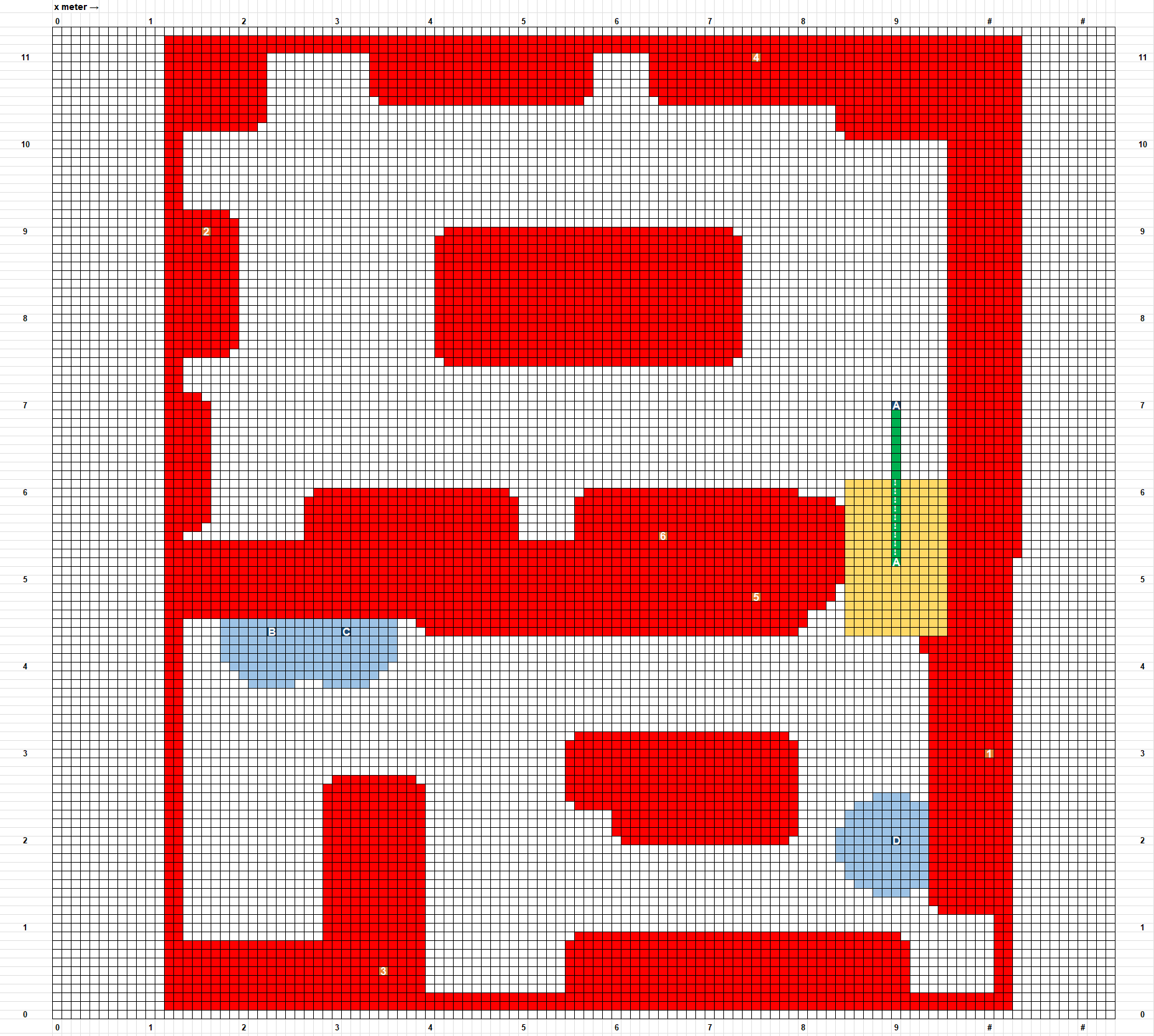}
\caption{Site-aware map guidance for the script-template example.  The planned
robot motion is shown against the DemoRoom geometry and restriction zones.  In
this example, a planned normal move enters the marked bump/ramp zone; the same
condition is reported in Fig.~\ref{fig:authoring-interface}(a) as
\texttt{NORMAL\_MOVE\_ENTERS\_RAMP\_ZONE}.  The pair illustrates the separation
between visual authoring guidance and authoritative Python validation.}
\label{fig:authoring-map}
\end{figure}

The principal authoring views are illustrated in
Figs.~\ref{fig:authoring-interface} and~\ref{fig:authoring-map}.  The captured
example also shows why the map is useful before submission: a planned normal move enters the marked
bump/ramp zone, which is visually apparent in the site-aware map and is then
reported explicitly by the Python validation result in the command sheet.

\begin{sidewaysfigure}[p]
\centering
\includegraphics[width=0.96\textheight,trim=0 0 0 45,clip]{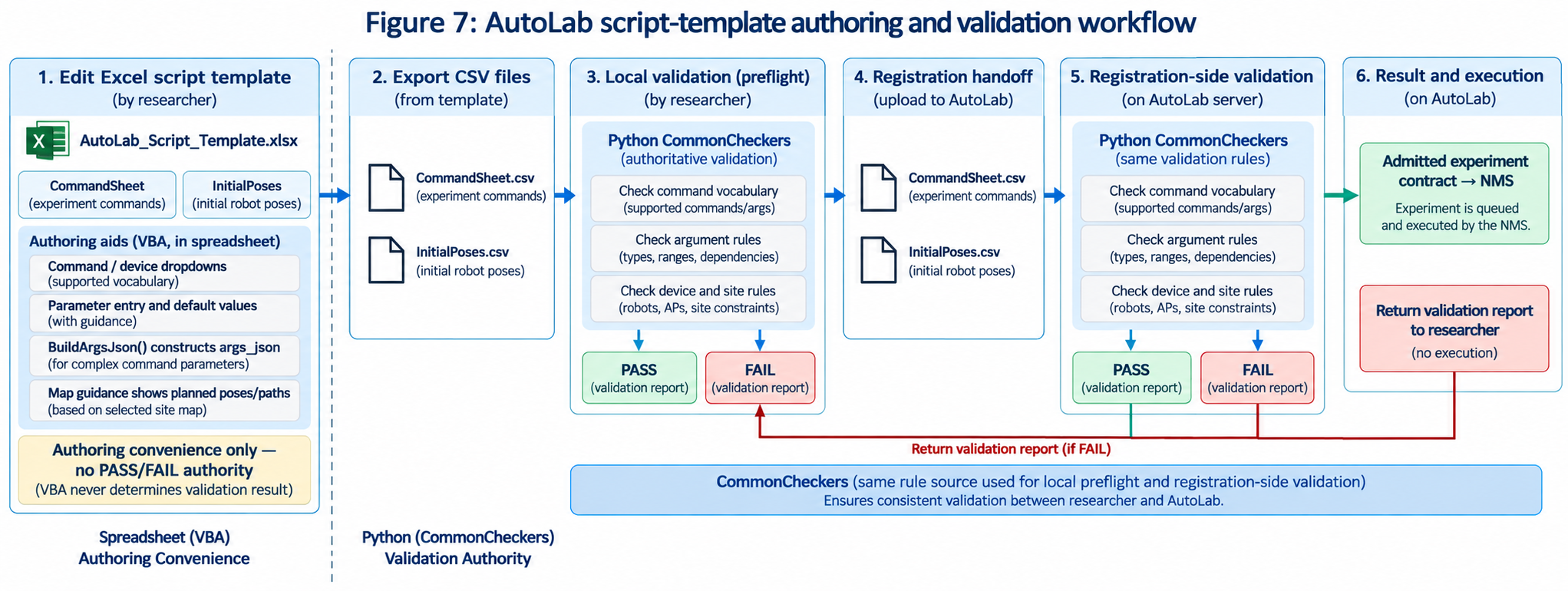}
\caption{AutoLab script-template authoring and validation workflow.  VBA assists
the researcher in constructing \texttt{CommandSheet}, \texttt{InitialPoses},
and \texttt{args\_json}, but it has no PASS/FAIL authority.  The exported CSV
contract is checked by the Python \texttt{CommonCheckers} during local
preflight and again at the AutoLab registration boundary using the same rule
source.  A passing contract is admitted for NMS execution; a failing contract
returns a validation report without execution.}
\label{fig:script-validation-workflow}
\end{sidewaysfigure}

Figure~\ref{fig:script-validation-workflow} summarizes the boundary between
spreadsheet authoring convenience and authoritative Python validation, including
the repeated use of the same \texttt{CommonCheckers} rule source before and
after registration handoff.

\clearpage
\clearpage

Once the exported contract passes registration, responsibility crosses the
boundary established in Section~\ref{sec:architecture}.  The Internet-side
authoring tool has described the intended scenario and demonstrated that it
satisfies the public validation contract; the NMS-side system then owns
registration, scheduling, command delivery, physical supervision, and runtime
records.  The author does not directly teleoperate motors, invoke arbitrary AP
software, or bypass site policy.  This separation is important both for
reproducibility and for safely exposing a physical laboratory to researchers
who are not present at the site.

The public workflow therefore presents one continuous researcher-facing path:
a researcher downloads the script template, describes the experiment, validates
the exported contract locally, submits the same contract to AutoLab for
authoritative revalidation and registration, observes the admitted experiment
through the LIVE interface, checks its recent execution through HISTORY, and
retrieves the experiment results through the service interface.  Site policy and
physical supervision remain behind the registration boundary, so extending
Internet access does not require exposing low-level robot or AP control to the
researcher.  Together, the authoring, observation, history, and result paths make
the public interface an operational entry point to the physical AutoLab rather
than merely a remote dashboard.

\part{NMS and Physical-Site View}

\section{Physical Test Site and NMS Architecture}
\label{sec:nms}

The Internet-facing workflow in Part~I ends when a validated experiment is
handed to the physical site.  From that boundary onward, the NMS is responsible
for turning the experiment contract into coordinated physical activity.  It
maintains the site-level view, schedules commands, communicates with device
agents, records their progress, and exports the resulting state and
measurements to the supporting services.  The researcher therefore interacts
with one experimental system rather than separately operating robots, APs, and
traffic processes.

\subsection{DemoRoom Physical Environment}

The first \autolab{} deployment is an indoor site identified as
\texttt{DemoRoom}.  Its site map uses a common metric coordinate system for
fixed infrastructure, mobile robots, planned destinations, and reported robot
poses.  The room contains inner and outer operating areas separated by a
doorway transition whose floor bump requires a controlled crossing procedure.
Ordinary destination-level movement is restricted from crossing this region;
dedicated inward-to-outward and outward-to-inward operations allow the NMS and
robot agent to apply the corresponding physical procedure explicitly.

The current DemoRoom configuration defines ten mobile-robot identities and six
logical AP identities.  A particular experiment may use only a subset of the
available devices.  AP locations are fixed by the site configuration, whereas
the initial robot poses are supplied as part of the experiment contract.  Each
robot combines mobility, camera-based AprilTag observation, wireless scanning,
and traffic-generation functions.  The robots use their onboard cameras to
capture raw observations of the site's AprilTag references; the NMS combines
these reports with the site map to estimate pose during movement.  A separate
Wi-Fi interface supports wireless observation and experiment traffic.
The APs expose a deliberately bounded set of management actions and status
reports rather than unrestricted remote administration.

\begin{figure}[p]
  \centering
  \begin{subfigure}[t]{0.30\textwidth}
    \centering
    \includegraphics[height=0.40\textheight]{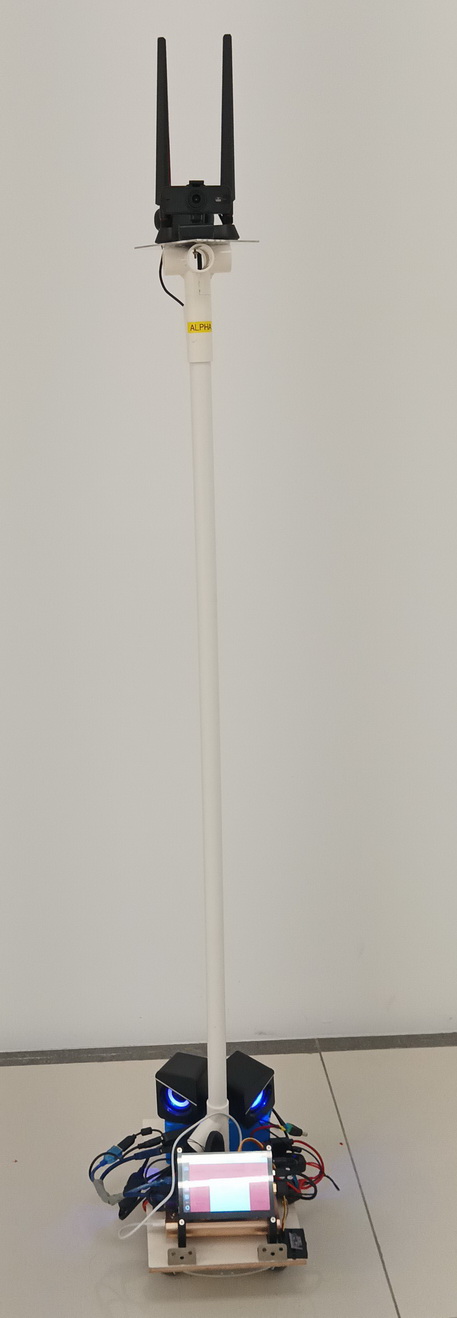}
    \caption{Complete mobile platform.}
  \end{subfigure}

  \vspace{0.7em}

  \begin{subfigure}[t]{0.47\textwidth}
    \centering
    \includegraphics[width=\linewidth]{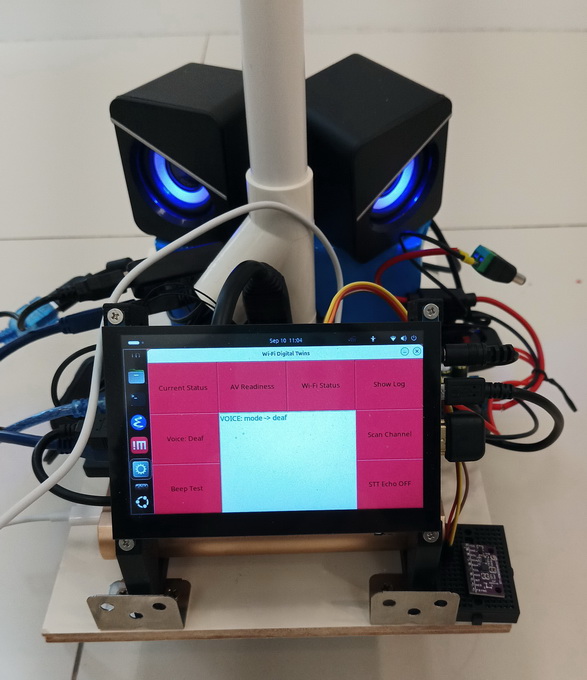}
    \caption{Front view of the mobile base.}
  \end{subfigure}
  \hfill
  \begin{subfigure}[t]{0.47\textwidth}
    \centering
    \includegraphics[width=\linewidth]{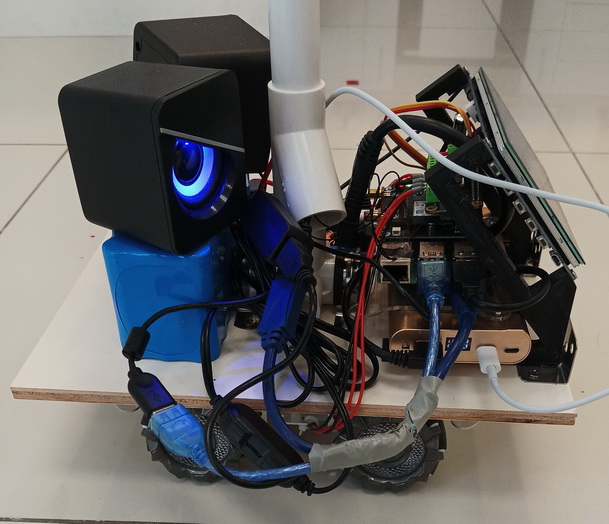}
    \caption{Side view of onboard hardware.}
  \end{subfigure}
  \caption{A mobile robot in the current \autolab{} deployment.  The elevated
  assembly supports observation and wireless experimentation, while the base
  integrates mobility, computation, power, device interfaces, and local status
  presentation.}
  \label{fig:robotplatform}
\end{figure}

Figure~\ref{fig:robotplatform} shows one of the mobile robots used in the
current deployment.  The complete view in Fig.~\ref{fig:robotplatform}(a)
shows the mobile base together with its elevated mast and radio/camera
assembly.  The closer front and side views in
Figs.~\ref{fig:robotplatform}(b) and~\ref{fig:robotplatform}(c) show that
mobility, onboard computation, user-visible status, power distribution, and
wired device interfaces are integrated into one physical platform.  This open
prototype construction also keeps the hardware interfaces accessible during
laboratory bring-up, diagnosis, and later extension.

The physical layout is not merely a background image for the web interface.
It is an executable part of the site definition.  Device positions, allowed
operating regions, restricted transitions, and site-specific rules are used
during experiment authoring, validation, execution, and later interpretation
of the measurements.  This shared site context is what allows a command stated
in physical coordinates to retain the same meaning from the public template to
the robot's measured outcome.

\begin{figure}[p]
  \centering
  \begin{subfigure}[t]{0.48\textwidth}
    \centering
    \includegraphics[width=\linewidth]{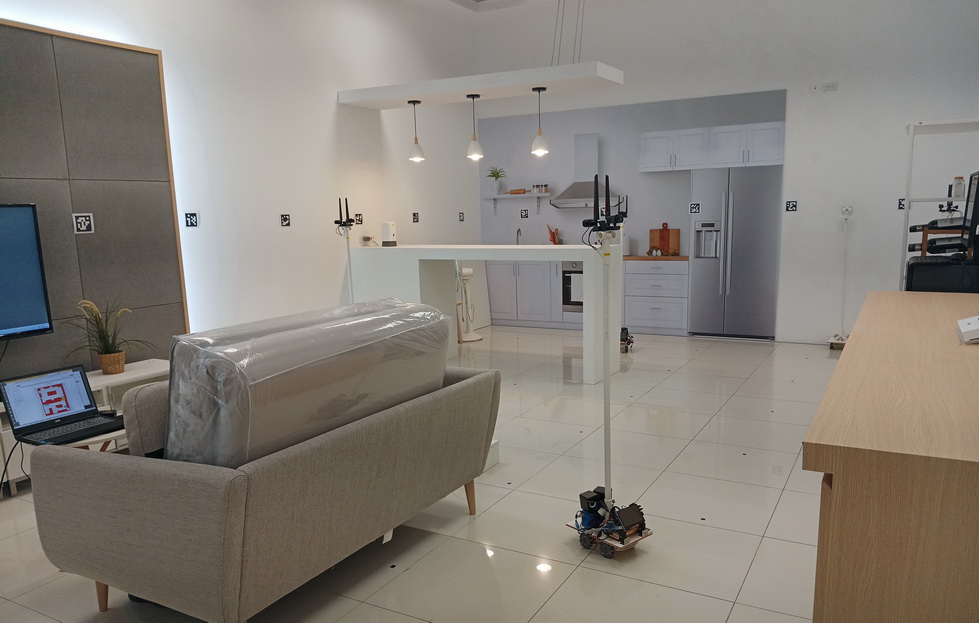}
    \caption{Lower room viewed from the bump area.}
  \end{subfigure}
  \hfill
  \begin{subfigure}[t]{0.48\textwidth}
    \centering
    \includegraphics[width=\linewidth]{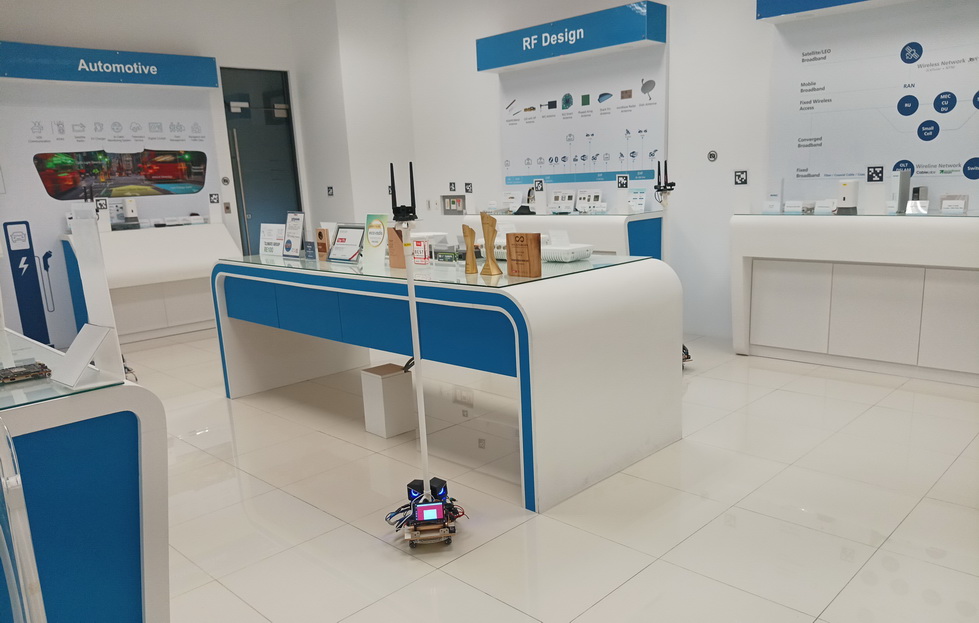}
    \caption{Upper room viewed from the bump area.}
  \end{subfigure}

  \vspace{0.7em}

  \begin{subfigure}[t]{0.66\textwidth}
    \centering
    \includegraphics[width=\linewidth]{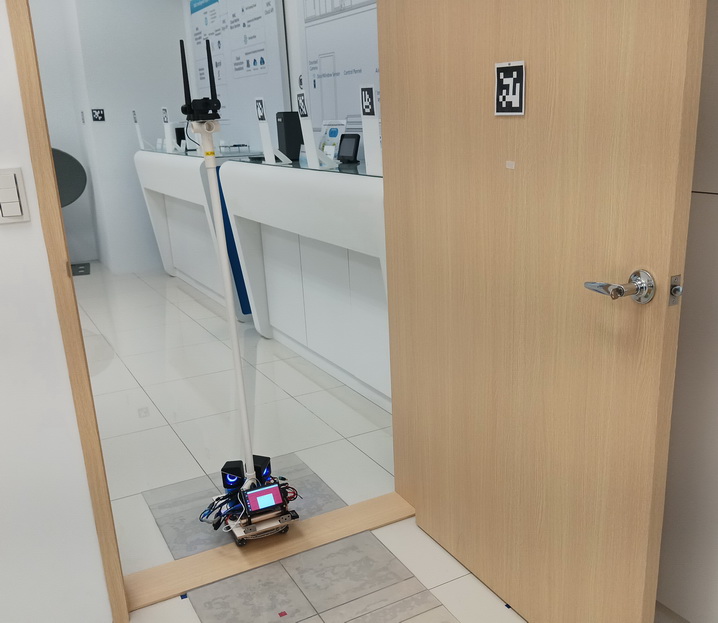}
    \caption{Doorway bump viewed from the inner room.}
  \end{subfigure}
  \caption{Physical views of the DemoRoom operating environment.  The room
  views expose robots, fixed wireless infrastructure, furnishings, and
  AprilTag references, while the doorway view shows the floor transition that
  requires a dedicated crossing procedure.}
  \label{fig:demoroomphotos}
\end{figure}

Figure~\ref{fig:demoroomphotos} provides three complementary physical views of
DemoRoom.  Figures~\ref{fig:demoroomphotos}(a) and
\ref{fig:demoroomphotos}(b) were taken from the doorway bump area while
looking into the lower and upper rooms, respectively.  They show the scale and
diversity of the operating environment, including furniture, fixed wireless
infrastructure, mobile robots, and distributed AprilTag references.  Figure
\ref{fig:demoroomphotos}(c) looks toward the same doorway transition from the
inner room.  The raised threshold is small relative to the room but significant
relative to the robot wheels; its physical form explains why ordinary
destination movement must not treat the two areas as one unobstructed plane.

\subsection{NMS-Centered Distributed Architecture}

The system uses centralized experiment coordination with distributed physical
execution.  The NMS holds the authoritative site schedule and experiment
state, but it does not directly drive motors or reproduce device-specific AP
commands.  Robot and AP agents run with the corresponding hardware and convert
the admitted high-level commands into local operations.  This separation keeps
global timing, policy, and bookkeeping at the NMS while leaving hardware
interaction and immediate execution checks at the device.

Table~\ref{tab:nmscomponents} summarizes the main physical-side components and
their responsibilities.  These are functional boundaries; several functions
may execute on the same host, and a robot agent may invoke multiple local
services while completing one command.

\begin{table}[t]
\centering
\caption{Principal components of the physical-side \autolab{} architecture.}
\label{tab:nmscomponents}
\begin{tabular}{p{0.22\textwidth}p{0.68\textwidth}}
\toprule
Component & Principal responsibility \\
\midrule
NMS & Registers experiments, maintains the site schedule, releases due
commands, collects device reports, and constructs the site-level operational
state. \\
Robot agent & Reports raw AprilTag observations, executes NMS-issued turns and
movements, performs wireless scans and robot-side traffic operations, and
reports progress, measurements, and failures. \\
AP agent & Applies the supported AP operations, reports radio/interface and
association state, and shields the NMS from device-specific management details.
\\
Mobility and location functions & Coordinate raw tag observation and physical
motion at the robot with site-map-based pose estimation, motion planning,
evaluation, and correction at the NMS. \\
Traffic execution & Creates the admitted TCP or UDP sessions at their scheduled
times and returns configuration, progress, and result information. \\
Web-service interface & Receives NMS snapshots and experiment information for
LIVE/HISTORY presentation and result access without obtaining direct authority
over physical actuators. \\
\bottomrule
\end{tabular}
\end{table}

An experiment command is addressed to a logical device and expressed at the
level required by the experiment.  For example, a mobility command supplies a
destination pose rather than individual wheel speeds, and an AP command names
a permitted operation rather than an arbitrary shell instruction.  The NMS
releases the command to the responsible agent, while the agent performs the
local sequence necessary to execute it.  Status and outcome reports then flow
back to the NMS, which relates them to the originating experiment and command.
Section~\ref{sec:execution} describes this supervised execution in detail.

This architecture also establishes a firm control boundary.  The website and
authoring tools describe, validate, and observe experiments, but only the NMS
admits scheduled work into the physical site.  Device agents accept work
through the NMS-facing protocol rather than directly from an Internet
researcher.  Consequently, improving the public interface does not require
exposing the robots' motor controls or the APs' native management interfaces.

\subsection{Device Identity and Site Configuration}

Physical hostnames are inconvenient experiment identifiers: they may be long,
device-specific, or changed during maintenance.  \autolab{} therefore
separates physical identity from the logical names used by experiment authors.
Robot registration associates the robot's control-interface hardware identity
with a configured logical name such as \texttt{twin-scout-alpha}.  The logical
name remains the target used in experiment scripts and NMS records.  This
allows hardware discovery and public experiment readability to coexist without
making a script depend on a transient network address.

AP identity has an additional layer because changing an AP's native hostname
is not always practical.  The site roster maps the native AP identity to a
short NMS-side key such as \texttt{AP1}; an optional application alias can be
used by the web service when a different presentation name is desirable.  The
experiment contract uses the stable logical AP key.  Thus, replacement or
renaming at one layer need not propagate into every public script, NMS record,
and display.

The site configuration binds these identities to the facts and policies of a
particular laboratory.  It includes the device rosters, fixed AP locations,
site-map assets, coordinate conventions, and restrictions needed by the public
checker and the NMS.  Site-independent command semantics remain in the common
validation package, whereas DemoRoom geometry and policy remain in the
DemoRoom-specific assets.  This organization permits another site to retain
the experiment contract and common command meanings while supplying its own
layout, roster, and admissibility rules.

\subsection{Command Polling, Reporting, and Service Boundaries}

Communication between the NMS and physical agents is periodic and
state-oriented.  In the current implementation, robot and AP agents use a
nominal ten-second polling cycle to request available work and report their
state.  The NMS uses these interactions to determine device availability,
associate command progress with the active experiment, and build the latest
site snapshot.  A separate periodic path forwards the relevant NMS state to
the web-side services, from which the LIVE and HISTORY views in
Sections~\ref{sec:live} and~\ref{sec:history} are constructed.

Polling provides a simple boundary between site coordination and heterogeneous
devices.  A temporarily unreachable agent does not require the website or
experiment author to understand the device-specific failure; the NMS observes
the missing or failed report, preserves the operational state, and applies the
appropriate experiment-level handling.  Conversely, a device report can carry
more than a binary completion flag.  Depending on the operation, it can include
pose estimates, scan observations, AP/interface state, association information,
traffic status or results, and diagnostic records.

The polling interval defines the nominal coordination cadence, not a claim
that every physical operation finishes within one interval.  Mobility, scans,
traffic sessions, and AP actions have different durations and completion
conditions.  The NMS therefore tracks their state across successive reports,
while the local agent performs the continuous or multi-step physical work.
This distinction is essential for long-running mobility and traffic operations:
scheduled release, physical completion, and result preservation are separate
events.

Taken together, the DemoRoom site definition, stable logical identities,
central NMS, and distributed agents form the physical execution substrate of
\autolab{}.  The next section follows a validated CSV contract as it is
rechecked, registered, assigned an absolute start time, and converted into the
schedule consumed by this architecture.

\section{Validation, Registration, and Scheduling}
\label{sec:validation}

Section~\ref{sec:authoring} described validation from the script writer's
viewpoint.  At the physical-site boundary, the same information has a stronger
meaning: it is a request to reserve a real laboratory and initiate physical
actions.  Registration therefore does not trust either the spreadsheet or a
previous PASS result.  It parses the submitted contract, selects the intended
site rules, repeats authoritative validation, and creates an executable
schedule only when the complete submission is accepted.

\subsection{Registered Experiment Contract and Identity}

The registration input consists of the two exported tables introduced in
Section~\ref{sec:architecture}.  \texttt{CommandSheet\allowbreak.csv} describes enabled,
time-ordered operations, their logical target devices, and their JSON
arguments.  \texttt{InitialPoses\allowbreak.csv} describes the starting poses needed to
interpret those operations at the selected site.  Neither file is meaningful
in isolation: command validity can depend on the initial pose, site geometry,
device role, and the relationship between multiple commands.

The laboratory operator submits the pair through the NMS registration path,
implemented in the current interface by
\texttt{POST /cmd/\_load\_csv\_file}.  The request also supplies the context
needed to bind the relative script to one physical execution, including the
laboratory identity and scheduled starting time.  The laboratory identity
selects the corresponding site configuration---\texttt{DemoRoom} in the
present deployment---rather than allowing the uploaded files to redefine the
physical site.

Each accepted submission is associated with an experiment identity used to
join its schedule, commands, reports, and results.  In the current workflow,
the uploaded command-file name supplies the basis for this identity, so the
operator must avoid registering ambiguous or conflicting names.  The logical
device names inside the contract are resolved through the selected site's
rosters; network addresses and native device hostnames are not embedded as
experiment targets.

Registration records the declared initial condition but does not imply that
the NMS has physically moved every robot to that pose.  The initial-pose table
states the condition under which the script is intended to begin.  Establishing
or confirming that physical condition remains part of preparing the site for
the scheduled run.  This distinction prevents experiment metadata from being
mistaken for a completed physical action.

\subsection{Authoritative Registration Validation}

The NMS invokes the same Python \texttt{CommonCheckers} implementation used by
the public preflight.  The files are parsed again from the submitted bytes, so
a local PASS result cannot authorize a file that was subsequently edited.
Likewise, registration does not reproduce a second set of validation rules in
the API handler.  The handler supplies the files and selected site context to
the common checker and uses its result to decide whether registration may
proceed.

Table~\ref{tab:registrationchecks} groups the principal admission checks.  The
categories are evaluated as one contract because an individually valid command
can still be invalid in relation to its target, time, initial condition, or
site.

\begin{table}[t]
\centering
\caption{Principal validation categories at the registration boundary.}
\label{tab:registrationchecks}
\begin{tabular}{p{0.24\textwidth}p{0.66\textwidth}}
\toprule
Category & Examples of enforced meaning \\
\midrule
File and schema integrity & Required tables and columns are present, values can
be parsed, and command arguments are valid JSON objects. \\
Command semantics & The command is publicly supported and its required,
optional, and prohibited arguments satisfy the common rules. \\
Device and role binding & The target exists in the selected site's roster and
is eligible for the requested robot, AP, scan, or traffic operation. \\
Temporal consistency & Enabled commands use admissible offsets and satisfy
cross-command timing restrictions. \\
Initial conditions & Required robot poses are present and consistent with the
devices and movements used by the experiment. \\
Site policy and geometry & DemoRoom operating bounds, restricted regions,
movement limits, and specialized crossing requirements are respected. \\
\bottomrule
\end{tabular}
\end{table}

Validation failure returns a structured report identifying the rejected rules
and prevents the experiment from entering the executable schedule.  A PASS is
therefore an admission decision, not merely an annotation attached to the
files.  This all-or-nothing boundary is important because partially registering
a multi-device experiment could leave commands without their intended initial
conditions or timing relationships.

The public preflight and registration check serve different trust boundaries
even though they execute the same rules.  Preflight provides rapid feedback to
the researcher before submission.  Registration protects the laboratory using
the exact bytes and site identity presented for execution.  Figure
\ref{fig:script-validation-workflow} shows these two invocations and their
shared rule source.

Admission-time validation should not be confused with runtime supervision.  A
valid script can still encounter wheel slip, an unavailable device, changing
radio conditions, or another physical condition that cannot be established
from CSV files.  Registration checks whether the requested experiment is
admissible; the device agents and NMS state machines determine whether its
physical actions are progressing safely.  Section~\ref{sec:execution}
describes that second protection layer.

\subsection{Absolute Start Time and Relative Command Time}

An experiment has one absolute starting time, denoted by $t_0$.  Each enabled
command row carries an offset $\Delta t_i$ relative to that start.  Its intended
release time is therefore
\begin{equation}
  t_i = t_0 + \Delta t_i.
  \label{eq:commandtime}
\end{equation}
This representation separates the structure of the experimental scenario from
the reservation of a particular laboratory time.  The same relative sequence
can be reviewed, validated, archived, and later scheduled at a different
$t_0$ without rewriting every command timestamp.

Relative time also makes the relationships among heterogeneous operations
explicit.  A traffic session can begin before a robot moves, an AP setting can
change between two scans, and a later measurement can be interpreted relative
to those interventions.  These offsets are part of the reproducible scenario,
whereas $t_0$ binds that scenario to a particular physical run.

The scheduled time is the intended release time, not a claim of zero-latency
physical completion.  Agents obtain work through the nominal polling cycle
described in Section~\ref{sec:nms}; operating-system delay, communication, and
the duration of a physical action separate command release from observed
completion.  \autolab{} therefore preserves scheduled time, execution state,
and reported outcome as distinct information.

\subsection{Schedule Construction and Command Delivery}

After a contract passes validation, the registration service associates its
enabled rows with the experiment identity, resolves their logical targets, and
stores their absolute release times.  Disabled authoring rows do not become
executable work.  The resulting schedule retains the command arguments and
relative ordering that were validated; registration does not silently rewrite
the requested experiment into a different scenario.

Robot and AP agents periodically identify themselves to the NMS and request
available commands.  The NMS compares the current time with the registered
schedule and releases due work only to the corresponding logical device.  A
robot therefore receives robot-addressed mobility, scan, or traffic work,
while an AP receives only admitted AP operations.  The local agent translates
the high-level command into the device-specific procedure and reports its
state and outcome through subsequent exchanges.

Command delivery and command success are deliberately separate.  Releasing a
command establishes that the scheduled work has crossed from the NMS to the
responsible agent; it does not assert that the requested physical result has
already been achieved.  Long-running actions remain observable across later
polling cycles, and failures remain associated with the originating command
and experiment rather than disappearing as transport errors.

This separation completes the transition from an Internet-authored script to
controlled physical execution:
\begin{equation}
\begin{split}
\text{submitted contract}
&\xrightarrow{\text{CommonCheckers}}
\text{registered experiment}\\
&\xrightarrow{\,t_i=t_0+\Delta t_i\,}
\text{device-addressed scheduled work}.
\end{split}
\label{eq:registrationflow}
\end{equation}
The next section follows that work inside the robot and AP agents, where
physical progress, correction, safety, and failure handling can no longer be
decided from the static contract alone.

\section{Physical Execution and Runtime Supervision}
\label{sec:execution}

Registration establishes that an experiment is admissible, but it cannot
guarantee the outcome of a physical action.  A wheel may slip, a camera may see
too few usable tags, a radio interface may be temporarily unavailable, or a
traffic process may terminate unexpectedly.  Runtime execution must therefore
observe progress and measured outcomes rather than equating command delivery
with success.  In \autolab{}, the NMS coordinates the experiment-level state,
while the robot and AP agents supervise the hardware-specific procedures that
run with their devices.

\subsection{Agent-Level Execution Model}

The commands admitted by Section~\ref{sec:validation} describe operational
intentions.  They do not expose low-level actuator sequences.  A destination
pose, for example, is realized through a coordinated NMS--robot procedure of
raw observation, centralized location estimation, motion planning, turning,
forward motion, post-motion evaluation, and possible correction.  Likewise, a
scan command controls a measurement process, an AP command invokes a bounded
management operation, and a traffic command creates a session with a declared
protocol and duration.

Each running operation is represented as a stateful procedure rather than as
one indivisible remote call.  This allows a command to remain observable
while physical work continues across multiple NMS polling intervals.  State
transitions record which phase is active, whether its preconditions were
available, what result was observed, and whether execution should advance,
retry, correct, stop, or fail.  The general runtime loop is
\begin{equation}
\begin{split}
\text{scheduled command}
&\rightarrow \text{local state machine}
\rightarrow \text{hardware action}\\
&\rightarrow \text{measured outcome}
\rightarrow \{\text{correct},\text{complete},\text{fail}\}.
\end{split}
\label{eq:runtime-loop}
\end{equation}

Table~\ref{tab:runtimecommands} summarizes how this pattern is instantiated by
the supported command families.  The shared pattern is supervision and
reporting; the completion evidence remains specific to the physical operation.

\begin{table}[t]
\centering
\caption{Runtime interpretation of supported experiment commands.}
\label{tab:runtimecommands}
\begin{tabular}{p{0.20\textwidth}p{0.34\textwidth}p{0.36\textwidth}}
\toprule
Command family & Supervised execution & Evidence returned to the NMS \\
\midrule
Mobility & Robot reports raw tags and executes NMS-issued turns and movements;
NMS estimates pose, evaluates error, and plans corrections & Raw tag reports,
estimated poses, residual errors, trace, and failure detail \\
Wireless scan & Perform one scan or manage a repeated scanning process & Scan
state, timestamped observations, and errors \\
AP operation & Translate an admitted logical operation into the AP's native
management mechanism & Applied operation, AP/interface state, association
state, and errors \\
Traffic session & Create the admitted TCP or UDP process and supervise its
declared lifetime & Session state, configuration, traffic results, and errors
\\
\bottomrule
\end{tabular}
\end{table}

\subsection{Destination-Level Robot Mobility}

The public command \texttt{mobility.move} specifies a desired planar pose
\begin{equation}
  \mathbf{q}_{\mathrm{d}} =
  \begin{bmatrix}x_{\mathrm{d}} & y_{\mathrm{d}} &
  \psi_{\mathrm{d}}\end{bmatrix}^{\mathsf T},
  \label{eq:desiredpose}
\end{equation}
where the final heading $\psi_{\mathrm{d}}$ may be omitted when only a
destination position is required.  The researcher does not specify wheel
speed, motor duration, or the individual turns used to reach this pose.  These
are determined through the coordinated NMS--robot mobility procedure.

Before translating the destination into motion, the robot captures raw
AprilTag observations and reports them to the NMS.  The NMS combines these
observations with the site map and registered tag locations to estimate the
current pose, compares it with the planned state, and computes the signed turn
angle and forward distance toward the destination.  The robot executes only
these physical instructions; it does not maintain the site map or calculate
its own global location and heading.  After motion, the robot captures and
reports another set of raw tag observations, from which the NMS estimates a
new pose.  For an achieved pose
$\hat{\mathbf{q}}=[\hat{x},\hat{y},\hat{\psi}]^{\mathsf T}$, the NMS
evaluates position and heading residuals such as
\begin{equation}
  e_p=\sqrt{(x_{\mathrm{d}}-\hat{x})^2+
             (y_{\mathrm{d}}-\hat{y})^2},
  \qquad
  e_\psi=\operatorname{wrap}
          (\psi_{\mathrm{d}}-\hat{\psi}).
  \label{eq:poseerror}
\end{equation}
When the applicable residual exceeds its configured tolerance, the NMS can
plan a corrective turn and movement from the measured pose rather than
continuing from an assumed trajectory.  Completion is declared from the
evaluated result, not solely from elapsed motor time.

This feedback is particularly important for the current robots, which are
constructed from low-cost components and operate on a real tiled floor.
Differences among motors, battery condition, wheel contact, and accumulated
turning error cannot be removed completely by one open-loop calibration.  The
distributed procedure instead combines calibration with repeated physical observation,
allowing moderate execution error to be absorbed through evaluation and
correction while still exposing persistent or excessive error as a failure.

Normal movement is bounded at both admission and runtime.  The current public
rules limit a normal destination move to at most three metres and require
moving commands to be separated sufficiently in the script; the present
DemoRoom rule uses a global separation of 180 seconds.  These limits reserve
time for location estimation, movement, evaluation, and correction, and reduce
the possibility that a later scheduled move overtakes an unfinished physical
procedure.  The 180-second separation is intentionally conservative for the
present implementation because a move can require several sensing, NMS
calculation, polling, physical-action, and correction cycles.  It can be
relaxed in a future site policy if accumulated execution evidence supports a
shorter interval without compromising supervision or command separation.

\subsection{Camera-Based Location Estimation}

Robot mobility is grounded by fixed AprilTag references distributed through
DemoRoom, as visible in Fig.~\ref{fig:demoroomphotos}.  Front and rear cameras
observe tag identity and relative geometry, but the robot does not interpret
these measurements in site coordinates.  It passes the raw tag information to
the NMS, where the registered tag locations, camera geometry, and site map are
combined to estimate planar position and heading.

Keeping the site map and location solver at the NMS is deliberate.  It avoids
replicating global site knowledge and calculation at every edge device,
maintains one coherent interpretation of tag locations among all robots, and
places robot locations in the same global state used for multi-robot
coordination and collision control.  An individual robot therefore need not
know where it is in the site coordinate system.  It observes the local physical
world, executes the turn angle and travel distance instructed by the NMS, and
returns new raw observations for the next centralized estimate.

The implemented NMS estimator uses a staged robust procedure.  Raw tag-wise
measurements first undergo basic screening.  A preliminary pose is then solved
without using yaw measurements.  This pose provides the reference for an
$M_2$ distance-consistency screen and for admitting and signing usable yaw
information.  The final component-wise solver incorporates the accepted
distance and yaw components to estimate the robot pose.  Separating preliminary
position recovery from yaw admission prevents a questionable orientation
measurement from determining its own interpretation.

Location estimation can fail legitimately when too few geometrically useful
tags are visible or when the surviving observations are inconsistent.  The
mobility procedure therefore treats ``no reliable pose'' as an explicit
runtime outcome rather than silently substituting the planned pose.  Depending
on the execution phase and configured policy, the NMS can request another raw
observation or instruct the robot to stop before further motion; the command
can also terminate with a diagnostic record.  Raw and screened observations,
intermediate decisions, and the resulting pose information can be retained for
later analysis of rare cases.

\begin{figure}[p]
  \centering
  \includegraphics[width=\textwidth,height=0.84\textheight,keepaspectratio]{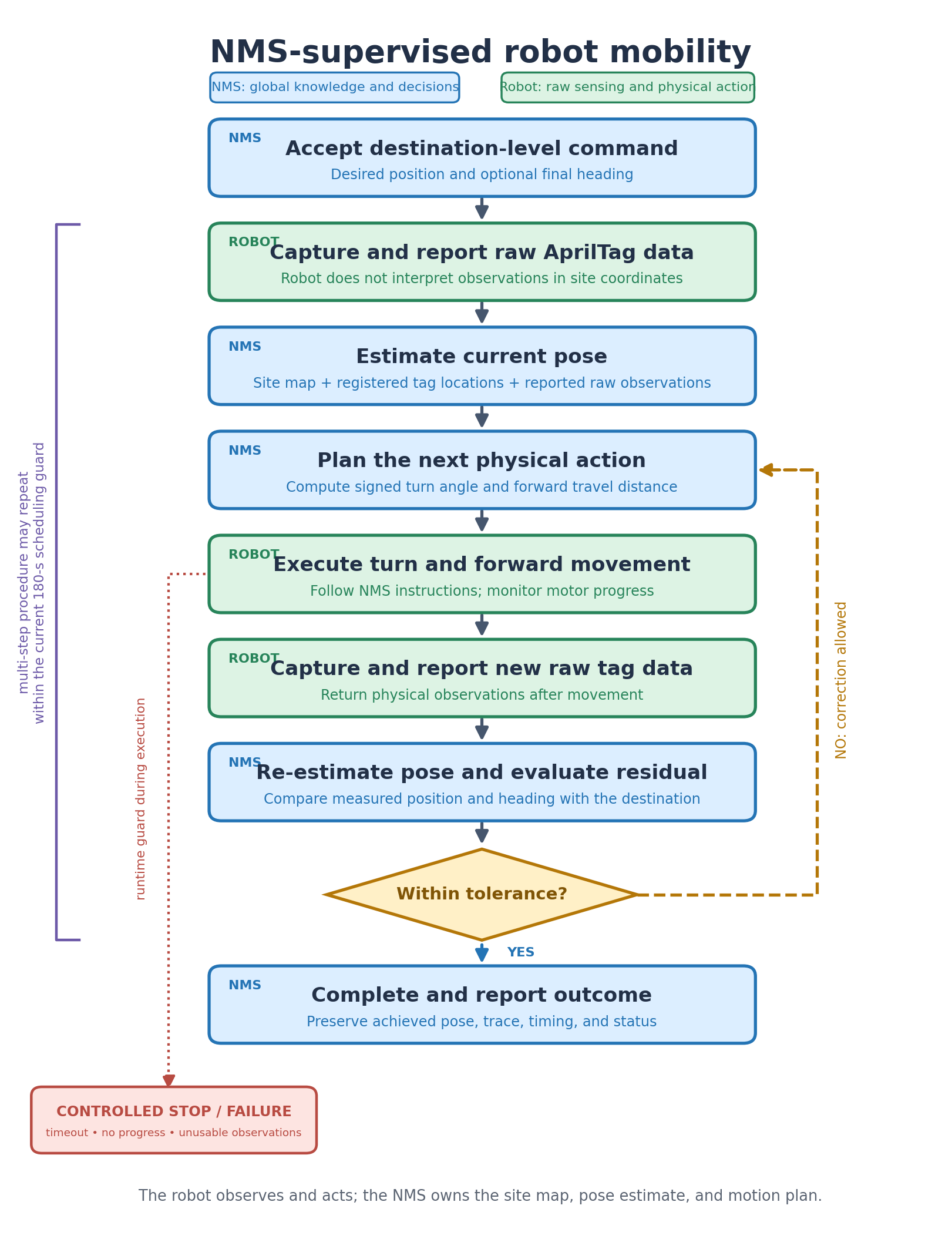}
  \caption{NMS-supervised robot mobility.  The robot supplies raw AprilTag
  observations and performs physical turning and movement; it does not store
  the site map or calculate its global pose.  The NMS combines the observations
  with registered site knowledge, estimates location and heading, plans the
  next physical action, evaluates the achieved result, and may issue a
  correction.  Runtime guards provide a controlled failure exit.  The bracket
  indicates why the current scheduling rule conservatively reserves 180
  seconds between mobility commands.}
  \label{fig:mobility-supervision}
\end{figure}

Figure~\ref{fig:mobility-supervision} summarizes this division of labor and the
major mobility states.  The long sequence between initial observation and
completion, together with the possible correction loop and polling delays,
also explains the conservative 180-second scheduling guard used at present.

\subsection{Controlled Doorway-Bump Crossing}

The doorway threshold shown in Fig.~\ref{fig:demoroomphotos}(c) violates the
assumption of an unobstructed planar move.  A normal
\texttt{mobility.move} is therefore not allowed to cross the configured bump
region.  The public vocabulary instead provides the direction-specific
commands \texttt{mobility.in2out} and \texttt{mobility.out2in}.  Making the
transition explicit allows validation to distinguish an ordinary destination
from a requested crossing and allows the NMS to select the corresponding
controlled physical procedure.

The specialized procedure preserves the same supervision principle as normal
mobility: the NMS issues the direction-specific motion instructions, the robot
performs the physical crossing and reports new raw tag observations, and the
NMS estimates the post-crossing pose and determines completion or failure.  The dedicated command also makes
crossing events identifiable in the experiment history, rather than hiding
them inside a geometrically similar destination request.  The commands and
their execution path are implemented.

\subsection{Wireless Scanning and AP Operations}

The robot command set provides \texttt{scan.once}, \texttt{scan.start}, and
\texttt{scan.stop}.  A one-shot scan produces one wireless observation at the
current physical condition, whereas start and stop delimit a repeated scanning
interval.  These commands currently require no public arguments.  The robot
agent selects the experiment Wi-Fi interface, controls the local scanning
process, and returns scan state and observations through the reporting path.
Because scanning is scheduled in the same script as mobility, AP changes, and
traffic, its measurements retain a defined temporal relationship to those
interventions.

AP commands follow the same high-level principle but execute in the AP agent.
The present public operations include bounded transmit-power control and
station disassociation.  The logical AP target is resolved through the site
roster, and the agent translates the admitted request into the AP's native
management operation.  The public contract does not expose an arbitrary AP
command channel.  This bounded interface allows experiments to modify selected
wireless-infrastructure conditions while preserving a known validation and
reporting path.

An AP operation is not considered physically meaningful merely because its API
call returned.  The agent reports the operation result together with subsequent
AP/interface or association state when available.  This distinction supports
experiments that compare an intended infrastructure change with the state and
traffic consequences actually observed afterward.

\subsection{TCP and UDP Traffic Sessions}

Traffic commands allow an experiment to create TCP or UDP sessions in the same
timeline as robot and AP activity.  A session has a script-visible identity,
target robot, duration, access category, and protocol-specific parameters.  The
robot agent translates the admitted description into the corresponding local
traffic process and reports session state and results.  Sessions normally run
for their declared duration, so a separate public stop command is not required
for the ordinary case.

Combining traffic and physical actions in one schedule is essential to the
experimental meaning.  For example, a session can establish offered load
before a robot moves, continue while association or received power changes,
and end after an AP intervention or scan.  The NMS preserves these timing
relationships while the agents supervise the different underlying processes.
Traffic is consequently part of the scenario being imposed on the physical
world rather than an unrelated measurement performed afterward.

\subsection{Runtime Safety, Failure Handling, and Records}

Runtime protection is layered on top of the admission rules.  Before and
during an action, the responsible agent checks the preconditions and progress
that can only be observed at the hardware.  Mobility supervision includes
bounded motion, phase timeouts, progress checks, controlled motor stopping,
post-motion pose evaluation, and bounded correction.  Scan, AP, and traffic
handlers similarly distinguish successful initiation from continued operation
and final outcome.

A failed phase does not justify continuing blindly to the next physical step.
If turning or forward motion makes insufficient progress, location estimation
does not provide a usable result, or a local process reports an error, the
agent stops the applicable activity and records the failure context.  The NMS
can then expose the command state without claiming that the requested outcome
was achieved.  Later commands remain governed by experiment-level policy and
the current device state rather than by an assumption that every preceding
action succeeded.

This design treats failure information as part of the experiment result.  A
mobility trace, rejected pose estimate, timeout, AP error, or traffic-process
termination can reveal physical limitations that an offline model would not
predict.  Preserving such events allows the same infrastructure that supervises
execution to contribute difficult cases to the operational history discussed
in Section~\ref{sec:records}.  The result is not fault-free automation, but a
controlled and observable interaction with imperfect physical hardware.

\section{Measurement, Reporting, and Experiment Records}
\label{sec:records}

Physical execution becomes reusable experimental evidence only when the
requested actions, observed conditions, and resulting outcomes remain related
after the run.  \autolab{} therefore treats reporting and record construction
as part of experiment execution rather than as an independent logging utility.
Robot and AP agents report measurements and progress to the NMS; the NMS binds
them to the site, logical device, experiment, and scheduled command, and then
constructs the current site state and the experiment-scoped records consumed by
the web and analysis services.  This path connects the physical-side execution
described in Section~\ref{sec:execution} with the LIVE and HISTORY views in
Sections~\ref{sec:live} and~\ref{sec:history}.

\subsection{Time- and Experiment-Referenced Reporting}

The nominal ten-second agent polling cycle provides repeated opportunities for
a device to receive work and return its current state.  A report may indicate
that a command has been accepted, that a multi-step physical procedure is still
in progress, that a measurement has been obtained, or that the procedure has
completed or failed.  Consequently, command release, physical observation, and
completion are distinct events; recording only the scheduled command time
would lose the behavior of the real system between them.

Conceptually, an operational record can be written as
\begin{equation}
  \mathcal{R}_k =
  \bigl(\tau_k,\,\ell_k,\,e_k,\,d_k,\,c_k,\,s_k,\,\mathbf{z}_k\bigr),
  \label{eq:operationalrecord}
\end{equation}
where $\tau_k$ is the observation or reporting time, $\ell_k$ identifies the
laboratory, $e_k$ the experiment, $d_k$ the logical device, $c_k$ the related
command or activity, $s_k$ its execution state, and $\mathbf{z}_k$ the
operation-specific payload.  Equation~\ref{eq:operationalrecord} is a
conceptual organization rather than a requirement that every device return the
same fields.  For a mobility operation, $\mathbf{z}_k$ may contain raw tag
observations, an NMS-estimated pose, or a motion residual; for a scan it may
contain timestamped radio observations; and for a traffic operation it may
contain session progress or results.

This common context is what makes heterogeneous reports jointly interpretable.
A received-power observation without a robot pose and time has limited value
for constructing a radio map.  A pose without the command and experiment that
produced it cannot explain whether it represents an initial condition, an
intermediate correction, or a completed destination.  By retaining identifiers
and timing alongside the operation-specific payload, the NMS can relate a
physical observation to the scenario that caused it.

\subsection{Measurements, State, and Diagnostic Records}

Table~\ref{tab:recordfamilies} summarizes the principal record families.  They
include both successful measurements and the evidence needed to interpret
incomplete or unsuccessful execution.  The latter is especially important in
a physical laboratory, where a timeout, inconsistent tag set, interrupted
traffic process, or unavailable AP state may reveal an operational condition
that was absent from the planned scenario.

\begin{table}[t]
\centering
\caption{Principal AutoLab record families and their experimental roles.}
\label{tab:recordfamilies}
\begin{tabular}{p{0.19\textwidth}p{0.34\textwidth}p{0.37\textwidth}}
\toprule
Record family & Representative contents & Experimental role \\
\midrule
Command state & Scheduled and released activity, target device, progress,
completion, and failure state & Relates intended actions to their observed
execution \\
Mobility and pose & Raw AprilTag reports, NMS-estimated poses, planned and
achieved motion, residuals, and mobility traces & Preserves spatial context and
the feedback path used to reach a destination \\
Wireless scan & Timestamped observations of visible wireless infrastructure
and received signal conditions & Supplies location-referenced radio evidence
when joined with the corresponding robot state \\
AP and station state & AP/interface condition, association information,
supported configuration changes, and operation results & Records the
infrastructure condition under which radio and traffic outcomes occurred \\
Traffic session & Protocol and admitted configuration, session state, timing,
results, and errors & Describes offered activity and its measured outcome \\
NMS site state & Active experiment, logical devices, poses, associations,
traffic activity, and command status & Provides a coherent site-level snapshot
for observation and historical reconstruction \\
Diagnostics & Timeouts, unusable observations, rejected estimates, local
process errors, and controlled stops & Preserves difficult physical cases
rather than silently treating them as missing data \\
\bottomrule
\end{tabular}
\end{table}

Not all entries are produced at the same rate or by the same component.  Raw
sensor and process information originates at the device, while global pose,
command association, and site-level state are constructed at the NMS.  This
division preserves the architecture of Section~\ref{sec:nms}: the robot reports
what it observes and does, whereas the NMS supplies the shared map, experiment
context, and interpretation needed across devices.  Derived information does
not replace its physical evidence.  For example, the estimated robot pose is
useful for the experiment and website, while retained raw tag information and
solver decisions remain useful for diagnosing a rare estimation failure.

\subsection{Current State, Short-Term History, and Experiment Records}

The reporting path supports three related but different uses of the same
physical evidence.  First, the NMS constructs the latest coherent site snapshot
for operational awareness.  This snapshot summarizes the active experiment,
devices, locations, associations, traffic sessions, and command states and is
forwarded to the web-side services for the LIVE view.  It represents what the
system most recently knows; it is not a claim that every field was measured at
exactly the same instant.

Second, the web-side history service records the periodically supplied state
stream for interactive reconstruction.  As described in
Section~\ref{sec:history}, its present 24-hour window allows a researcher to
move through recent site states and inspect spatial and operational transitions
without manually aligning independent device logs.  This history is optimized
for near-term observation and debugging.  It is distinct from physically
executing the experiment again and from the experiment-result path used for
subsequent analysis.

Third, experiment-scoped results preserve the measurements and runtime records
needed to analyze an admitted run.  Experiment identity and command identity
provide the bookkeeping spine: command states, mobility traces, scans, AP and
station observations, traffic results, and diagnostics can be traced back to
the script and initial condition that produced them.  This separation permits
the LIVE snapshot to remain compact and the HISTORY interface to remain
responsive without reducing a completed experiment to the fields currently
shown on the dashboard.

Long-term preservation is intentionally flexible rather than defined as one
exhaustive, permanent copy of every available field.  The measurements,
sampling density, intermediate results, and derived quantities retained for a
dataset can be selected according to the intended model-training or analysis
task.  A radio-map dataset, for example, emphasizes timestamped received-power
observations joined with robot poses, whereas a mobility study may preserve raw
tag observations, solver decisions, instructed motion, achieved poses, and
correction traces.  In either case, the selected data retain the experiment,
device, time, command, and site context needed to interpret and reproduce their
meaning.  This permits future datasets to evolve with the OWM algorithms
without requiring the operational platform to predict one final training
schema in advance.

The three uses can therefore be summarized as
\begin{equation}
  \text{device reports}
  \rightarrow \text{NMS operational state}
  \rightarrow
  \begin{cases}
    \text{latest snapshot for LIVE},\\
    \text{recent state stream for HISTORY},\\
    \text{experiment-scoped records for analysis}.
  \end{cases}
  \label{eq:recordpaths}
\end{equation}
They are different projections of one execution history, not three independent
accounts of what occurred.

\subsection{From Experiment Records to Operational Experience}

For OWM development, the important product is not merely a collection of
measurements but a sequence of interactions.  At time $k$, the model or
researcher has an observed operational context $\mathbf{o}_k$, applies or
schedules an action $\mathbf{a}_k$, and later observes an outcome
$\mathbf{y}_{k+1}$.  A completed experiment contributes experience of the form
\begin{equation}
  \mathcal{E} =
  \left\{(\mathbf{o}_k,\mathbf{a}_k,\mathbf{y}_{k+1})\right\}_{k=0}^{K-1},
  \label{eq:operationalexperience}
\end{equation}
with site, time, device, and experiment context supplied by the records above.
This organization supports questions that disconnected output files cannot
answer reliably: what radio and traffic state preceded an action, what physical
change was actually achieved, what service consequence followed, and whether
the outcome required correction or ended in failure.

Repeated experiments can accumulate such experience into the long-term prior
knowledge discussed in Section~\ref{sec:owm}.  A new run then need not represent
learning the world from an empty state.  It can use established knowledge of
radio behavior, device limitations, spatial structure, and successful or
unsuccessful interventions while absorbing the much smaller amount of new
evidence specific to the current site and scenario.  Records from unusual
conditions are valuable in this process because an OWM intended to endure
unknown futures must learn from cases outside a single fixed script, including
cases in which the intended action did not produce the expected state.

\autolab{} does not by itself prescribe one learning algorithm or require a
complete representation of the physical environment.  It supplies controlled
actions, machine-native observations, operational outcomes, and the
bookkeeping needed to relate them.  The next section evaluates this implemented
path through end-to-end demonstrations and concrete system evidence.

\section{Implementation and End-to-End Demonstration}
\label{sec:demonstration}

The preceding sections have already shown the physical site, robot platform,
public interfaces, experiment contract, and NMS architecture.  Repeating those
views here would demonstrate only that the components exist.  This section
instead examines what the integrated system does when physical execution
deviates from the plan.  We use an overnight mobility experiment to show two
complementary outcomes: automatic correction when the NMS obtains a reliable
post-motion pose, and controlled termination when the physical observations do
not support a reliable pose.

\subsection{Experiment and Evidence Source}

A submitted experiment describes a nominal twelve-hour run by robot
\texttt{twin-scout-bravo}.  Its 241 high-level rows
comprise one initial location report, 236 normal destination moves, and four
direction-specific doorway crossings.  Commands are spaced by 180 seconds.
The experiment begins at 19:21 and, if uninterrupted, ends with the final
crossing at 07:21.  The script therefore tests ordinary movement throughout
the inner and outer areas as well as repeated transitions across the doorway
bump.

The corresponding \texttt{mobility\_trace} records the NMS-side events generated
as commands are converted into physical actions.  These events include the
computed command, its issuance, the robot report received by the NMS, the pose
solution, the residual evaluation, any correction command, and the resulting
state transition.  The robot report also carries its local command-received
and command-finished timestamps.  Because NMS and robot timestamps originate
from different clocks, the analysis uses NMS timestamps for ordering
inter-system events and uses the robot timestamps only to report local
execution duration; it does not infer sub-second communication latency.

The run reached the scheduled 01:21 crossing before entering a controlled stop.
Before that event, the trace contains 119 completed script-level movement
commands.  Of these, 110 were accepted after the first post-motion evaluation,
while nine caused the NMS to generate a correction and were subsequently
accepted after that correction.  These counts characterize this run only; they
are not presented as a general mobility-performance distribution.

\subsection{Automatic Absorption of a Physical Movement Error}

\begin{figure}[p]
  \centering
  \includegraphics[width=0.94\textwidth]{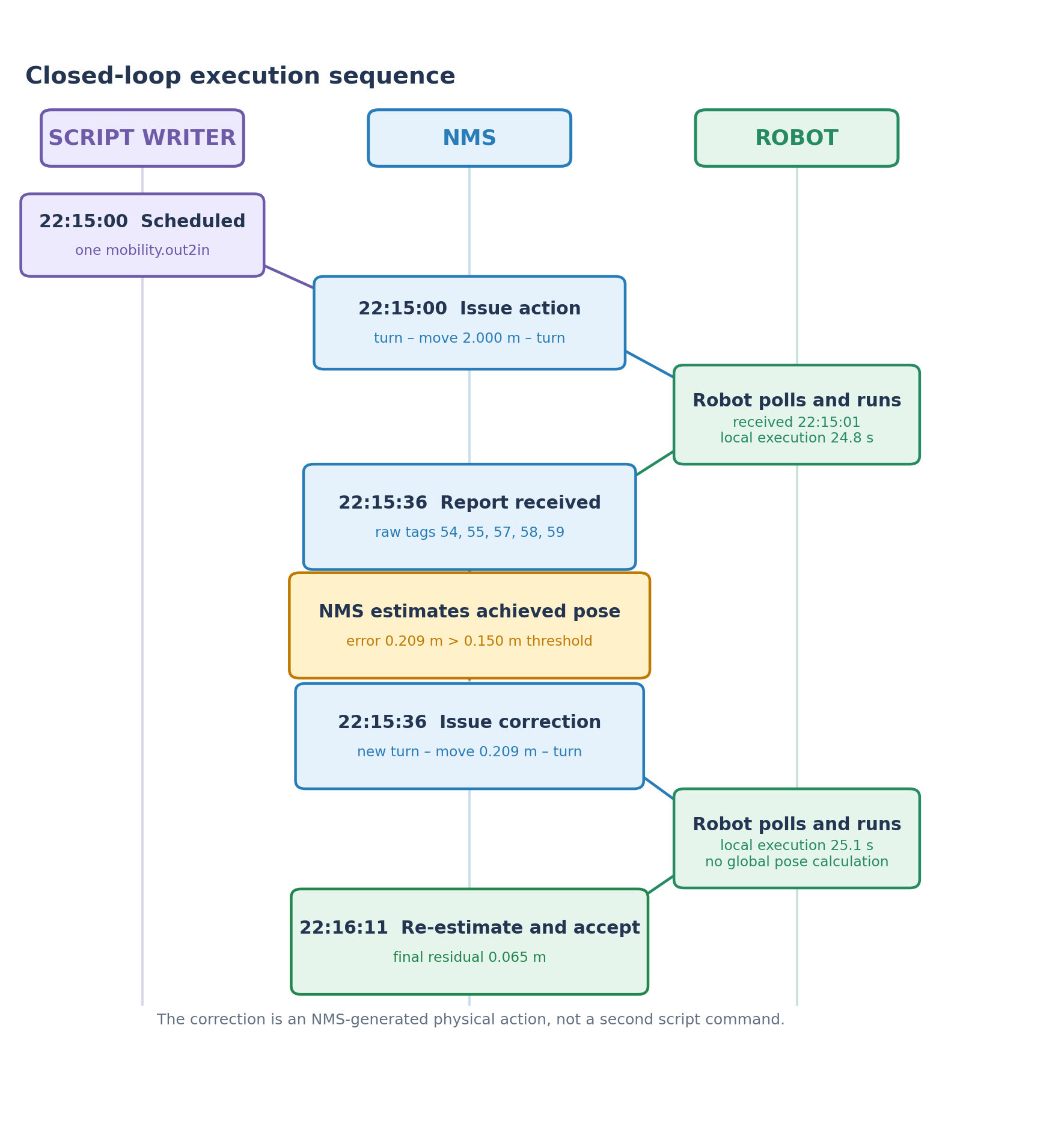}
  \caption{Closed-loop execution sequence for the 22:15
  \texttt{mobility.out2in} operation.  One public script command passes through
  NMS issuance, robot polling and local execution, raw-observation reporting,
  centralized pose estimation, an NMS-generated correction, and final
  acceptance.  NMS times order cross-system events; execution durations are
  computed from robot-local timestamps.  The correction is not a second
  command written by the script author.}
  \label{fig:correction-sequence}
\end{figure}

\begin{figure}[p]
  \centering
  \includegraphics[width=0.52\textwidth]{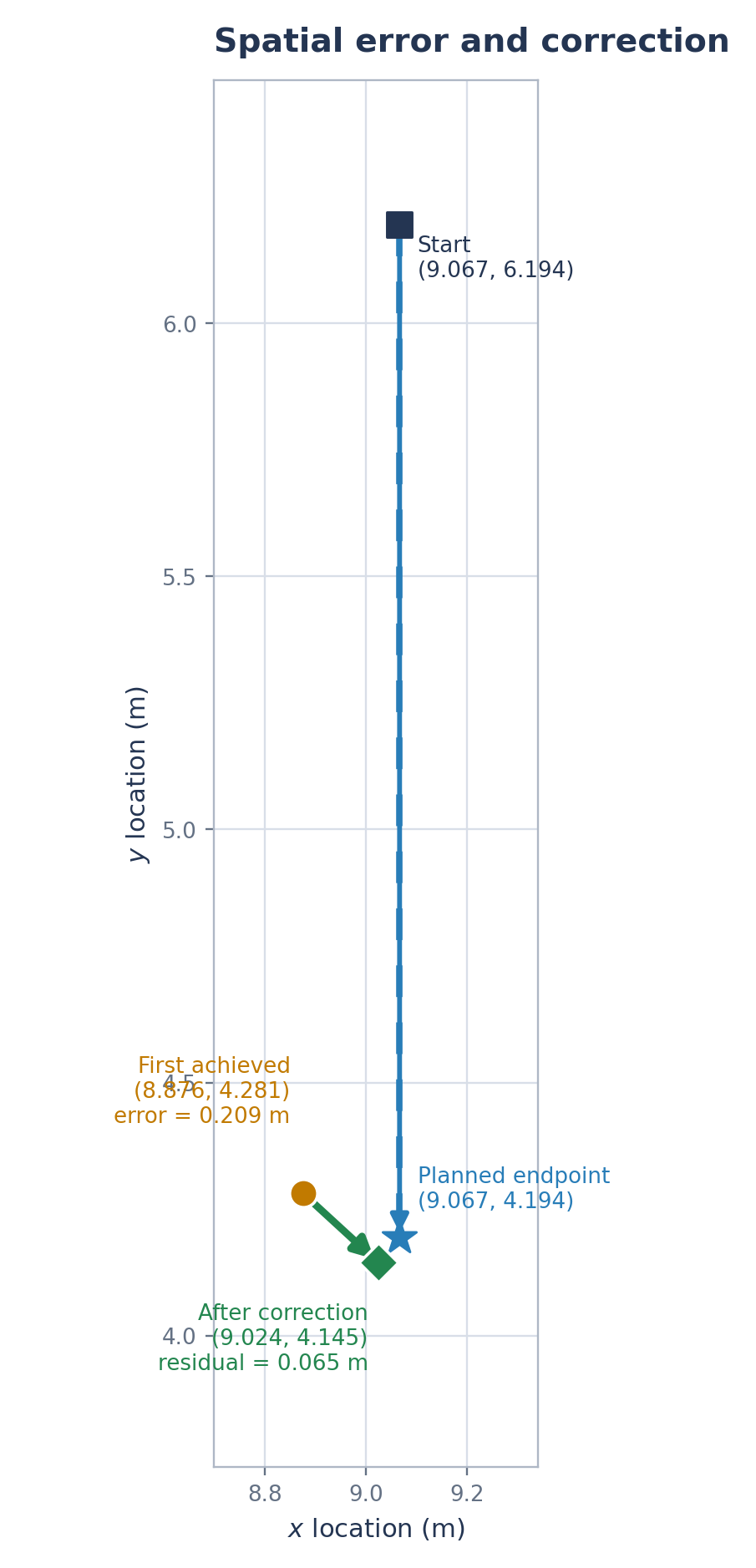}
  \caption{Spatial view of the same correction event.  The first physical
  movement ended $0.209$~m from the planned endpoint, exceeding the
  $0.150$~m correction threshold.  After the NMS-generated corrective motion,
  the estimated residual was reduced to $0.065$~m and accepted.}
  \label{fig:correction-geometry}
\end{figure}

Figures~\ref{fig:correction-sequence} and~\ref{fig:correction-geometry} expand the
\texttt{mobility.out2in} command scheduled at 22:15.  The public script contains
one high-level crossing request and no corrective command.  Using the measured
starting pose, the NMS generated the first physical
\texttt{turn--move--turn} action with a forward distance of $2.000$~m and
issued it at 22:15:00.  Bravo obtained the command through its polling path and
completed the local action in approximately $24.8$~s.  The robot did not
calculate a global pose; it returned execution status and raw observations of
tags 54, 55, 57, 58, and 59.

The NMS received this report at 22:15:36 and combined the observations with its
registered site knowledge.  The planned endpoint was
$(9.067,4.194)$~m, whereas the NMS-estimated achieved position was
$(8.876,4.281)$~m.  The resulting position residual was
\begin{equation}
  e_p = 0.209~\text{m} > e_{\mathrm{trigger}}=0.150~\text{m},
  \label{eq:democorrectiontrigger}
\end{equation}
and the pose solution had HIGH confidence.  The NMS therefore determined that
correction was justified and generated another
\texttt{turn--move--turn} action whose forward distance was $0.209$~m.

Bravo received this second physical instruction through the same agent
interface and completed it in approximately $25.1$~s.  A new raw-tag report
reached the NMS at 22:16:11.  From it, the NMS estimated the robot position as
$(9.024,4.145)$~m, leaving a final position residual of $0.065$~m.  The
mobility procedure then recorded the outcome as
\texttt{accepted\_after\_correction} and returned to its idle state before the
next scheduled script command.

The correction is internal to the physical execution of the original
high-level request.  The script writer does not predict the error or insert a
second row.  The robot does not decide that its global position is wrong; it
only reports raw observations and executes the next NMS instruction.  The
Internet researcher can continue to observe the experiment at the level of
the requested operation and resulting site state without manually controlling
the recovery.  Thus, the physical discrepancy is not ignored, but it is
absorbed below the public experiment contract.

\subsection{Controlled Termination When Correction Is Not Justified}

Automatic correction depends on adequate physical evidence.  At 01:21, the
script requested a \texttt{mobility.in2out} crossing.  The NMS generated and
issued the corresponding two-metre bump-crossing action, and Bravo reported
local completion.  Its post-motion camera result, however, contained only tags
25 and 26.  After the first screening layer, only tag 25 remained usable for
the required distance-and-angle calculation.  The NMS estimator consequently
returned no global position or heading, with the diagnostic
\texttt{insufficient Layer-1 usable distance+angle observations}.

The NMS did not substitute the planned endpoint for the missing measurement and
did not issue a position correction from an unreliable estimate.  It processed
the report into state \texttt{s7stopped}, preserving the observed tags, failed
pose result, command context, and stopping reason.  The overnight experiment
therefore ended before its nominal completion time, but it failed in a
controlled and interpretable manner.  This event demonstrates the other side
of supervision: moderate physical error can be corrected when the evidence is
sufficient, whereas lack of trustworthy state information prevents continued
blind motion.

\subsection{What the Demonstration Establishes}

The correction case is an end-to-end implementation result rather than an
offline replay of a motion model.  One script command caused NMS planning,
physical execution by a low-cost robot, raw environmental observation,
centralized location estimation, quantitative residual evaluation, an
unwritten corrective action, and a verified final outcome.  Across the portion
of this run completed before the stop, all nine script-level movements that
crossed the correction threshold were followed by NMS-generated correction and
an accepted post-correction result.

The controlled-stop case is equally important.  It shows that a failed
experiment need not become an unstructured absence of data.  The same trace
that reveals the failure also records the requested action, physical report,
surviving observations, estimator decision, and final execution state.  Such
cases can expose weak observation regions, changes in robot behavior, or
limitations of the current procedure and can be selected for later diagnostic
or model-training datasets according to the flexible preservation policy of
Section~\ref{sec:records}.

These examples do not establish a universal error rate, guarantee recovery
from every physical deviation, or demonstrate a completed autonomous OWM.
They establish the implemented capability required at the OWM grounding
boundary: \autolab{} can impose a scripted scenario, observe what actually
happened, compare outcome with intention, intervene below the public command
interface when justified, and preserve both corrected and uncorrectable cases
as structured operational experience.

\section{Related Work}
\label{sec:related}

\subsection{World Models and Operational World Models}

World-model research studies internal representations that allow an agent to
predict environmental evolution and evaluate actions before applying them.
Ha and Schmidhuber demonstrated that a generative model could learn a compact
spatial--temporal representation of a reinforcement-learning environment and
support a policy trained within imagined trajectories~\cite{ha2018world}.
LeCun subsequently described a broader architecture for autonomous machine
intelligence in which a configurable world model predicts abstract
representations of possible future states and supports planning
~\cite{lecun2022path}.  These works motivate prediction and action through an
internal model, but their principal examples concern visually observed or
embodied-agent environments.

The OWM proposed here is compatible with this broad direction while making a
different design commitment explicit.  Its state is selected by operational
sufficiency and may be formed primarily from machine-native observations such
as received-power fields, contention, traffic signatures, service margins, and
network configuration.  Human-visible reconstruction is optional rather than
defining.  The OWM also emphasizes accumulated long-term operational knowledge,
adaptation to the current episode, preventive intervention, and physical
feedback.  \autolab{} contributes the mechanism for obtaining that feedback;
it does not propose a competing general-purpose world-model learning
architecture.

\subsection{Digital Twins and Wireless Network Models}

Digital-Twin literature grew from the connection between a physical product or
system and its digital counterpart.  Grieves and Vickers described the Digital
Twin as a means to use information about physical and virtual systems to detect
and mitigate undesirable behavior~\cite{grieves2017digital}.  In wireless
networking, learned twins have been proposed to estimate performance indicators
under alternative network configurations more efficiently than repeated
simulation~\cite{li2023learnable}.  Colosseum has also been presented as an
Open-RAN Digital Twin: its hardware-in-the-loop channel emulator and
softwarized protocol stacks reproduce controlled RF and network scenarios for
repeatable development and testing~\cite{polese2024colosseum}.

These systems illustrate that replication, emulation, learning, and control can
all appear within Digital-Twin work.  Our distinction is therefore one of
system identity rather than a claim that every Digital Twin is passive.
\autolab{} interacts with an over-the-air physical Wi-Fi environment instead
of replacing its propagation by an RF channel emulator.  More importantly,
the surrounding research objective begins with accumulated experience and
current operational state, generates plausible futures, evaluates their risk,
and seeks preventive action.  The earlier predictive and preventive wireless
framework in~\cite{chen2026predictive} supplies this theoretical direction;
the OWM terminology avoids implying that faithful physical replication is its
endpoint.

\subsection{Remotely Accessible Wireless Testbeds and Emulators}

Large shared wireless facilities established the value of remotely accessible,
programmable experimentation.  POWDER provides a city-scale, end-to-end
software-defined platform with remote access and control across radio,
network, compute, and application layers~\cite{breen2021powder}.  Colosseum
provides a complementary form of scale and repeatability through
hardware-in-the-loop RF emulation~\cite{polese2024colosseum}.  Such platforms
enable experiments that would be difficult to reproduce using isolated local
equipment and provide important infrastructure for developing and testing
data-driven wireless control.

\autolab{} does not seek to reproduce their geographic scale, radio-node count,
or cellular protocol-stack scope.  Its contribution is a different integration
point: a physically furnished indoor Wi-Fi site in which mobile robots, AP
operations, wireless scans, and traffic sessions are combined in one
time-ordered public experiment contract.  The same contract passes through
author-side and registration-side validation, NMS scheduling, supervised
physical execution, live observation, historical reconstruction, and
dataset-oriented record preservation.  The centralized NMS also absorbs
moderate mobility error below the public command interface, as demonstrated in
Section~\ref{sec:demonstration}.  This complete Internet-to-physical lifecycle,
rather than testbed scale alone, is the focus of the platform.

\subsection{Robotic Wireless Measurement and Autonomous-Vehicle Testbeds}

Mobile robots and unmanned vehicles have been used both as wireless clients and
as measurement instruments.  Pandey and Parasuraman measured bidirectional
Wi-Fi performance for indoor mobile robots, including throughput, delay,
retransmissions, and signal strength while robots carried sensor traffic
~\cite{pandey2021wifi}.  Work combining Digital Twins with autonomous-vehicle
testbeds has compared simulation, software-in-the-loop,
hardware-in-the-loop, and physical experimentation, and has demonstrated
AI-assisted signal-source search using real link-quality observations
~\cite{gurses2024avntwins}.  These efforts show why mobility and real radio
measurements are important for evaluating algorithms outside a static network.

In \autolab{}, robotic measurement is one component of a broader operational
experiment.  The Internet researcher specifies destination-level movement and
its timing relationship to scans, AP changes, and traffic rather than
teleoperating the robot.  Robots report raw AprilTag information, while the NMS
holds the common site map, estimates global pose, coordinates devices, checks
residual error, and issues correction when justified.  The resulting record
therefore relates machine-native radio observations not only to robot movement
but also to the complete imposed scenario and its physical outcome.

Across these areas, individual ingredients of \autolab{} have strong
precedents: world models predict consequences, Digital Twins connect physical
and virtual systems, shared testbeds provide remote experimentation, and robots
collect spatial wireless evidence.  The contribution reported here is their
system-level combination into a validated, observable, and supervised physical
experience loop intended to ground wireless OWMs.

\section{Conclusion}
\label{sec:conclusion}

This paper has presented \autolab{} as an Internet-accessible physical
experimentation platform for wireless OWMs.  The platform joins functions that
are often treated separately: public observation of a live wireless site,
historical reconstruction, structured experiment authoring, common validation
at two trust boundaries, NMS registration and scheduling, supervised execution
by heterogeneous devices, and preservation of physical measurements and
runtime evidence.  The validated experiment contract connects the public and
physical-site views, allowing the researcher to specify what operational
scenario should occur without receiving unrestricted low-level control of the
laboratory hardware.

The implementation also illustrates why centralized physical supervision is
part of the experimental contribution rather than merely an internal service.
Robots report raw observations and execute bounded turn-and-move instructions;
the NMS retains the common site map, estimates pose, evaluates residual error,
and decides whether correction is justified.  In the demonstrated trace, this
division allowed a large movement error to be detected and corrected beneath
the destination-level experiment interface.  The resulting record preserves
the issued commands, physical observations, estimated state, correction, and
outcome as one spatial--temporal experience.

\autolab{} does not yet constitute a complete autonomous OWM, nor does this
paper claim general online scenario generation or fully deployed predictive
and preventive network optimization.  It provides the observation,
intervention, execution, feedback, and bookkeeping infrastructure upon which
those capabilities can be constructed and tested.  Continuing work will use
this interface to develop richer operational-state learning, generative
spatial--temporal scenarios, predictive risk screening, preventive network
management, and transfer of accumulated operational knowledge across
experiments and sites.

More broadly, \autolab{} records a physical experiment not as isolated output
files but as a controlled interaction between a proposed scenario and the real
world.  This makes it possible for researchers to design experiments remotely,
observe their execution, inspect unexpected behavior, retrieve task-specific
datasets, and evaluate models against repeatable physical evidence.  By opening
that interaction loop, \autolab{} supplies a practical path from machine-native
wireless observations to future OWMs that can learn from experience, anticipate
operational difficulty, and eventually act before service failure occurs.

\end{document}